\documentclass[draftcls,onecolumn,12pt]{IEEEtran}

\usepackage{cite}
\usepackage{bm}
\usepackage{url}
\usepackage[usenames,dvipsnames]{color}
\usepackage[dvipsnames]{xcolor}
\usepackage{amsmath}
\usepackage{mathtools}
\usepackage[]{amssymb}
\usepackage{amsthm}
\usepackage{float}
\usepackage{dsfont}
\usepackage{array}
\usepackage{booktabs}
\usepackage{multirow}
\usepackage{footmisc}

\usepackage{graphicx}
\usepackage[para,flushleft]{threeparttable} 
\usepackage{multirow}	
\usepackage{dcolumn} 

\usepackage[ruled]{algorithm2e}
\usepackage{xifthen} 
\usepackage{enumitem}

\makeatletter
\newcommand{\removelatexerror}{\let\@latex@error\@gobble}
\makeatother

\DeclareMathOperator{\diag}{diag}

\newif \ifwebcolor
\webcolortrue

\def \ba {\begin{array}}
\def \ea {\end{array}}
\def \benu {\begin{enumerate}}
\def \eenu {\end{enumerate}}
\def \bdes {\begin{description}}
\def \edes {\end{description}}

\def \bitem {\begin{itemize}}
\def \eitem {\end{itemize}}

\def \bfl {\begin{flushleft}}
\def \efl {\end{flushleft}}
\def \bfr {\begin{flushright}}
\def \efr {\end{flushright}}

\def \beq {\begin{equation}}
\def \eeq {\end{equation}}
\def \bqa {\begin{eqnarray}}
\def \eqa {\end{eqnarray}}
\def \bqa* {\begin{eqnarray*}}
\def \eqa* {\end{eqnarray*}}

\def \bal {\begin{align}}
\def \eal {\end{align}}

\newcounter{mytempeqncnt}

\DeclareMathOperator*{\argmax}{arg\,max}

\begin{document}
%
\title{Online Expectation-Maximization Based Frequency and Phase Consensus in Distributed Phased Arrays}

\author{\IEEEauthorblockN{Mohammed Rashid, ~\IEEEmembership{Member,~IEEE}, and 
Jeffrey A. Nanzer, ~\IEEEmembership{Senior Member,~IEEE}}

\thanks{Manuscript received 2022.}
\thanks{This work has been submitted to the IEEE for possible publication. Copyright may be transferred without notice, after which this version may no longer be accessible.}
\thanks{This work was supported in part by the Office of Naval Research under grant number N00014-20-1-2389. Any opinions, findings, and conclusions or recommendations expressed in this material are those of the author(s) and do not necessarily reflect the views of the Office of Naval Research. \textit{(Corresponding author: Jeffrey A. Nanzer.)}}
\thanks{The authors are with the Electrical and Computer Engineering Department, Michigan State University, East Lansing, MI 48824 (e-mail: \mbox{rashidm4@msu.edu,} nanzer@msu.edu).}}


\maketitle
\thispagestyle{empty}
\pagestyle{empty}

\def\bda{\mathbf{a}}
\def\bdd{\mathbf{d}}
\def\bde{\mathbf{e}}
\def\bdf{\mathbf{f}}
\def\bdg{\mathbf{g}} 
\def\bdh{\mathbf{h}}
\def\bdm{\mathbf{m}}
\def\bds{\mathbf{s}} 
\def\bdn{\mathbf{n}}
\def\bdu{\mathbf{u}}
\def\bdv{\mathbf{v}}
\def\bdw{\mathbf{w}} 
\def\bdx{\mathbf{x}} 
\def\bdy{\mathbf{y}} 
\def\bdz{\mathbf{z}}
\def\bdA{\mathbf{A}}
\def\bdC{\mathbf{C}}
\def\bdD{\mathbf{D}} 
\def\bdF{\mathbf{F}}
\def\bdG{\mathbf{G}} 
\def\bdH{\mathbf{H}}
\def\bdI{\mathbf{I}}
\def\bdJ{\mathbf{J}}
\def\bdU{\mathbf{U}}
\def\bdX{\mathbf{X}}
\def\bdK{\mathbf{K}}
\def\bdQ{\mathbf{Q}_n}
\def\bdR{\mathbf{R}}
\def\bdS{\mathbf{S}}
\def\bdV{\mathbf{V}}
\def\bdW{\mathbf{W}}
\def\bdGamma{\bm{\Gamma}}
\def\bdgamma{\bm{\gamma}}
\def\bdalpha{\bm{\alpha}}
\def\bdmu{\bm{\mu}}
\def\bdSigma{\bm{\Sigma}_n}
\def\bdxi{\bm{\xi}}
\def\bdl{\bm{\ell}}
\def\bdLambda{\bm{\Lambda}}
\def\bdeta{\bm{\eta}}
\def\bdPhi{\bm{\Phi}}
\def\bdpi{\bm{\pi}}
\def\bdtheta{\bm{\theta}}
\def\bdTheta{\bm{\Theta}}
\def\bddelta{\bm{\delta}}

\def\btau{\bm{\tau}}
\def\deg{\circ}

\def\tq{\tilde{q}}
\def\tbdJ{\tilde \bdJ}
\def\l{\ell}
\def\bdzero{\mathbf{0}} 
\def\bdone{\mathbf{1}} 
\def\Exp{\mathbb{E}} 
\def\exp{\text{exp}} 
\def\ra{\rightarrow}

\def\R{\mathbb{R}} 
\def\C{\mathbb{C}} 
\def\CN{\mathcal{CN}} 
\def\N{\mathcal{N}} 
\begin{abstract}
Distributed phased arrays are comprised of separate, smaller antenna systems that coordinate with each other to support coherent beamforming towards a destination. 
However, due to the frequency drift and phase jitter of the oscillators on each node, as well as 
the frequency and phase estimation errors induced at the nodes in the process of synchronization, 
there exists a decoherence that degrades the beamforming process.
To this end, a decentralized frequency and phase consensus (DFPC) algorithm was proposed in prior work for 
undirected networks in which the nodes locally share their frequencies and phases 
with their neighboring 
nodes to reach a synchronized state. Kalman filtering (KF) was also integrated with DFPC, and the resulting KF-DFPC 
showed significant reduction in the total residual phase error upon convergence. One limitation of DFPC-based 
algorithms is that, due to relying on the average consensus protocol which requires undirected networks, 
they do not converge for directed networks. Since directed networks can be more easily implemented in practice, 
we propose in this paper a push-sum based 
frequency and phase consensus ($\text{P}_\text{s}$FPC) algorithm which converges for such networks. 
The residual phase error of $\text{P}_\text{s}$FPC upon convergence is theoretically derived in this work. Kalman filtering 
is also integrated with $\text{P}_\text{s}$FPC and the resulting KF-$\text{P}_\text{s}$FPC algorithm 
shows significant reduction in the residual phase error upon convergence. Another limitation of KF-DFPC 
is that it assumes that the model parameters, i.e., the 
measurement noise and innovation noise covariance matrices, are known in KF. Since they may not be known in 
practice, we develop an online expectation maximization (EM) based algorithm 
that iteratively computes the maximum likelihood (ML) estimate of 
the unknown matrices in an online manner. EM is integrated with KF-$\text{P}_\text{s}$FPC and the resulting 
algorithm is referred to as the EM-KF-$\text{P}_\text{s}$FPC algorithm. Simulation results are included where 
the performance of the proposed $\text{P}_\text{s}$FPC-based algorithms is analyzed for different distributed phased arrays 
and is compared to the DFPC-based algorithms and the earlier proposed hybrid 
consensus on measurement and consensus on information (HCMCI) algorithm. 
\end{abstract}


\section{Introduction}\label{intro_section}

Modern advancements in wireless technologies 
have enabled more reliable communications between separate wireless systems, 
and as a result, several new applications involving distributed systems have emerged over recent years. 
These include distributed beamforming systems \cite{dbeam_2022}, 
 distributed automotive radars \cite{dautoradar_1, dautoradar_2}, 
and  distributed massive MIMO systems \cite{Distd_mMIMO}, among others. 
The underlying antenna array technology in all these applications is a distributed phased array (a.k.a.
distributed antenna arrays, virtual antenna arrays, or coherent distributed arrays). 
A distributed phased array is essentially a network 
of multiple spatially distributed and mutually independent antenna systems that are 
wirelessly coordinated at the wavelength level to coherently transmit and 
receive the radio frequency signals~\cite{Nanzer_Survey_2021}. Thus, compared to the classical analog or digital 
phased arrays where all the antennas are mounted together on a single-platform, coherently 
distributed phased array architectures brings in significantly better spatial diversity, higher directivity, improved signal 
to noise ratio at the destination, greater resistance to failures and interference, reconfigurable capabilities for the dynamic environments, 
and ease of scalability. 

In single-platform phased arrays, it is common that all antennas and their transceiver chains 
are driven by a universal local oscillator to assure the synchronized state of the array, 
whereas in distributed phased arrays, each antenna system has its own 
local oscillator in its transceiver chain. Thus, in a free-running state in which these oscillators are not locked 
to any reference signal, they undergo random frequency drift and phase jitter 
which results in decoherence of the signals emitted by the nodes. 
While reference signals from systems such as global positioning system (GPS) satellites can be used to 
synchronize these oscillators across the array 
provided that the nodes are equipped with the GPS receivers \cite{GPS_syn}, it is not always the case that GPS is available or reliable.
Alternatively, these nodes can also be synchronized by wirelessly transmitting the reference signals from the destination system; 
several such methods have been proposed that include single or multiple bit 
feedback methods~\cite{1_bit_feedback, 3_bit_feedback}, or  
retrodirective synchronization~\cite{retrodirective_2016}. However, since these approaches rely on the feedback from the 
target destination, they are essentially closed-loop, and cannot steer beams to arbitrary directions.  
These closed-loop methods are thus only suitable for 
the wireless communication applications where such feedback from the target destination is feasible. 
Recently, open-loop synchronization methods have also been proposed~\cite{OCDA_2017, rashid2022frequency, Serge_2021, dbeam_2022}, in which 
the nodes communicate with each other to synchronize their oscillators without using any 
feedback from the destination. Since the destination need not be an active system, open-loop distributed phased arrays can also be used for radar and remote sensing applications. 

In an open-loop distributed phased array, the electrical states (i.e., the frequencies and phases) of the nodes 
can be synchronized by either using a centralized approach or 
a decentralized approach. In \cite{Serge_2021, dbeam_2022, Mabel_2017, Mabel_2019}, 
a centralized open-loop distributed phased array was used in which  
nodes are classified as either a primary node or a secondary node. 
The primary node transmits a reference signal to the secondary ones which is used to synchronize their electrical 
states across the array. However, this architecture is susceptible to the primary node's failure, and 
given the limited communication resources, it is also not scalable. On the other 
hand, in decentralized (i.e., distributed) 
open-loop distributed phased array systems, there is no such classification of the nodes as either the primary or 
secondary nodes. Therein, the nodes communicate with their neighboring nodes and iteratively 
share their measurements to update their electrical state, 
until synchronization is achieved across the array. 
Based on this approach, a decentralized frequency and phase consensus (DFPC) algorithm was proposed in 
\cite{rashid2022frequency} wherein the nodes update these parameters in 
each iteration by computing a weighted average of the shared values using the average 
consensus algorithm \cite{Fast_MC_2004}. 
Kalman filtering was also integrated with the DFPC algorithm which further reduced 
the residual phase error upon convergence.
The synchronization and convergence performances of 
the KF-DFPC algorithm were also compared to the DFPC algorithm, 
the diffusion least mean square (DLMS) algorithm \cite{LMS_2010}, the 
Kalman consensus information filter (KCIF) algorithm \cite{Saber_2009}, and the diffusion Kalman 
filtering (DKF) algorithm \cite{Sayed_2010, Xin_2022}, 
where it was shown that KF-DFPC outperforms DFPC and DLMS in terms of reducing the residual phase 
error, and that for the same residual phase error, KF-DFPC 
converges faster that the DFPC, KCIF, and DKF algorithms which makes it more favorable when the latency in 
achieving the synchronized state is undesirable or when low-powered nodes are used in the distributed phased array.   

However, there are two major limitations of the above algorithms which are as follows. First, 
these algorithms require bidirectional (undirected) communication links between the nodes, whereas in practice due to the large spatial separation between the nodes, the dynamic changes in their environments, and their limited communication 
ranges, there may exist unidirectional (directed) communication links between them. The traditional average consensus algorithm \cite{Fast_MC_2004} does not converge for directed networks because the algorithm requires that the weighting 
matrix of the network 
be doubly-stochastic (i.e., both its rows and columns must sum to one), whereas for the directed 
networks, it is usually a column stochastic matrix. 
Secondly, the above algorithms, in particular, the KF-DFPC algorithm of \cite{rashid2022frequency}, 
assume that the statistical knowledge of the frequency drifts and phase jitters of 
the oscillators as well as that of the frequency and phase estimation errors are known \textit{a priori}. 
Specifically, it is assumed that 
the innovation noise and the measurement noise covariance matrices in the state transition model 
and the observation model for the Kalman filtering algorithm are known to the nodes, however in practice this is not always possible since
the operational dynamics of the oscillators are influenced by several factors, 
such as the changes in their operating temperatures and the 
power supply voltages \cite{ModPara_vary_1}, 
aging of the oscillators \cite{Filler_1993, Wang_2016}, and noise induced by their electronic components
\cite{ModPara_vary_2}. Furthermore, the statistics of the 
frequency and phase estimation errors are estimated from the observed estimates at the nodes where their accuracies 
are influenced by the signal to noise ratio (SNR)
of the observed signals and the numbers of samples collected per observation period. Thus, 
the noise covariance matrices are generally unknown to the nodes and must be estimated for Kalman filtering. 

A push-sum method \cite{PS_1} based decentralized 
frequency synchronization algorithm was proposed in \cite{Hassna_TAP} for 
the directed networks of distributed phased array. The authors assumed a specific class of the nodes in which the 
frequencies are only incremented in discrete frequency steps, which limits the frequency resolution of the nodes 
and increases the residual phase error of the algorithm. 
In this paper, we consider a more advanced class of nodes, e.g., USRP based nodes 
as used in \cite{dbeam_2022}, for which such frequency quantization errors are negligible. 
Furthermore, unlike the work in \cite{Hassna_TAP}, we consider both frequency and phase synchronization of 
the nodes in distributed phased array, both of which are necessary for distributed beamforming.
The contributions of this article are summarized as follows.  

\begin{itemize}
\item We consider the same signal model for the nodes as assumed in \cite{rashid2022frequency} 
in which the oscillators induced offset errors, such as 
the frequency drifts, the phase drifts, and the 
phase jitters are included and modeled using the practical statistics. 
Due to these added offsets, the exact frequencies and phases are unknown to the nodes and must estimated from the 
observed signals. Thus, we also include the frequency and phase estimation errors 
in our signal model. However, in this paper, we consider directed communication links between the nodes, i.e., if a node transmits an 
information to another node, it may not hear back from it. For such networks, we propose 
a push-sum algorithm based decentralized joint frequency and phase consensus algorithm 
for the directed networks in distributed phased array, which is 
referred to herein as the $\text{P}_\text{s}$FPC algorithm. Note that our proposed $\text{P}_\text{s}$FPC algorithm 
is not a direct extension of the push-sum based frequency synchronization algorithm of \cite{Hassna_TAP}, in fact 
it is a modified version where the estimated frequencies and phases of the nodes are included and has to be 
multiplied by the weighting factor 
from the previous iteration before computing the temporary variables (see Eqn. \eqref{temp_PsFPC} below). This step is 
required to recover the temporary variables from the previous iteration taking into account the fact that the updated 
frequency and phases from each iteration of $\text{P}_\text{s}$FPC 
undergo a state transition due to the random drifts and jitter in between the update intervals. 

\item The residual phase error of the $\text{P}_\text{s}$FPC algorithm 
is also theoretically derived to analyze the contributing factors for the residual decoherence between the nodes  
upon its convergence. It is observed that the phase error depends on the network algebraic connectivity (i.e., 
the modulus of the second largest eigenvalue $\lambda_2$ of the weighting matrix) which decreases with the increase 
in the number of nodes in distributed phased array or with the increase in the connectivity between the nodes. Simulation results are 
also included where the residual phase errors of $\text{P}_\text{s}$FPC are 
analyzed in the context of these theoretical results for different distributed phased array networks.
\item In distributed phased arrays, the electrical states of the nodes are updated with smaller update intervals, usually on the order 
of milli-second (ms), to 
avoid large oscillator drifts \cite{Serge_Access_2021} and to correct the residual phase errors due to the platform vibrations \cite{Pratik_2017}. 
However, it was shown in \cite{rashid2022frequency} that at such update intervals, 
the residual phase error of a consensus algorithm is usually higher 
when no filtering is used at 
the nodes, which deters an accurate coherent operation. 
Kalman filtering (KF) is a popular algorithm that provides the optimal minimum mean squared error (MMSE) 
estimates of the unknown quantities when their states transitioning model follows 
the first-order Markov process, and the observations are a linear function of the states. 
Thus, we integrate KF with $\text{P}_\text{s}$FPC to improve the residual 
phase errors for the directed networks of distributed phased array. 
Furthermore, in contrast to \cite{rashid2022frequency}, 
we assume in this work that the measurement noise and innovation 
noise covariance matrices are unknown for the KF, and thus 
we derive an online expectation-maximization (EM) algorithm \cite{Cappe_EM_2009} to compute their maximum likelihood (ML) 
estimates from the observed measurements. The online EM algorithm is integrated 
with KF and $\text{P}_\text{s}$FPC, and the resulting algorithm is referred to herein as the 
EM-KF-$\text{P}_\text{s}$FPC algorithm. 
The computational complexity analysis of EM-KF-$\text{P}_\text{s}$FPC is 
also included at the end of Section \ref{EM_section} 
where it is shown that our proposed online EM algorithm has the same computational complexity as that of KF, and 
notably for the moderately connected large arrays, using the EM and KF algorithms 
does not increase the overall computational 
complexity of the $\text{P}_\text{s}$FPC algorithm.
\item Simulations results are included where the 
residual phase errors and the convergence speeds of the proposed $\text{P}_\text{s}$FPC and 
EM-KF-$\text{P}_\text{s}$FPC algorithms are analyzed by varying the number of nodes in the array and the 
connectivity between the nodes. Furthermore, 
our online EM algorithm is also integrated with the KF-DFPC algorithm of \cite{rashid2022frequency} and the performance 
of the resulting EM-KF-DFPC is also compared to the EM-KF-$\text{P}_\text{s}$FPC algorithm 
and the HCMCI algorithm of \cite{HCMCI_2015}. 
\end{itemize}

Rest of this article is organized as follows. Section \ref{offset_modeling} describes the signal model of the 
nodes, the proposed $\text{P}_\text{s}$FPC algorithm, and theoretically analyzes the residual phase error 
of $\text{P}_\text{s}$FPC upon its convergence. 
Simulation results are also included therein to analyze the synchronization 
performance of $\text{P}_\text{s}$FPC. The KF algorithm and the 
online EM algorithm are derived in Section \ref{EM_section} where these algorithms 
are integrated with the $\text{P}_\text{s}$FPC algorithm to improve the synchronization between the nodes. 
The performance of EM-KF-$\text{P}_\text{s}$FPC is analyzed through simulations and is also compared to 
EM-KF-DFPC for the undirected networks. To this end, the initialization of both the EM and KF algorithms are 
also described in Section \ref{EM_section} and 
the computational complexity analysis of the overall algorithm is also performed. 

\textbf{{Notations}:} A lower case small letter $(x)$ is used to represent 
a signal or scalar, and a lower case bold small letter $(\bdx)$ is used to denote a vector. An 
upper case bold capital letter $(\bdX)$ represents a matrix.
The transpose and inverse operations on a matrix are indicated by the 
superscripts $(.)^T$ and $(.)^{-1}$, respectively. The notations 
$|\bdX|$ and tr$\{\bdX\}$ are used for the determinant and trace of a matrix $\bdX$, respectively. 
A normal probability distribution on 
a random vector $\bdx$ is denoted by $\N(\bdmu,\mathbf{\Sigma})$ in which $\bdmu$ indicates the 
mean and $\mathbf{\Sigma}$ represents the covariance matrix of the distribution. 
The expectation of $\bdx$ with respect to the 
probability distribution on $\bdx$ is represented by $\Exp[\bdx]$. 
For ease of notation, a set $\left\{\bdx_n(1),\bdx_n(2),\ldots,\bdx_n(k)\right\}$ 
is written in compact form as $\bdx^{1:k}_n$. $\bdone$ is a column vector of all $1$s. 
Finally, a diagonal matrix created with the elements in vector $\bdx$ is 
denoted by $\bdX=\diag\{\bdx\}$, and an $N\times N$ identity matrix is indicated by $\bdI_N$.


\section{Frequency and Phase Synchronization in Distributed Phased Array 
with Directed Communication Links}\label{offset_modeling}

Consider a distributed phased array made up of $N$ spatially distributed antenna nodes 
that are coordinating with each other 
to perform a coherent operation toward the destination. 
We assume that due to the large spatial separation between the nodes 
and their limited communication ranges, 
the communication links between the nodes are unidirectional across the entire array. 
Thus we model this array network by a directed graph 
$\mathcal{G}=(\mathcal{V},\mathcal{E})$ in which $\mathcal{V}=\{1,2,\ldots,N\}$ represents the set of vertices 
(antenna nodes), 
and $\mathcal{E}=\{(n,m)\colon n,m\in \mathcal{V}\}$ is the set of all directed edges (unidirectional 
communication links) in the array. Let the signal generated by the $n$-th node in 
the $k$-th iteration be given by $s_n(t)=e^{j\left(2\pi f_n(k) t+\theta_n(k)\right)}$ for $t\in[(k-1)T,kT]$, where $T$ is the signal duration, and 
$k\in\{1,2,\ldots\}$. 
The parameters $f_n(k)$ and 
$\theta_n(k)$ represent the frequency and phase of the $n$-th node in the $k$-th iteration, respectively.  
In a decentralized consensus averaging algorithm, the nodes iteratively exchange their frequencies and phases  
with their neighboring nodes and update these parameters by computing a weighted average of the shared values 
until convergence is achieved (i.e. these parameters are synchronized 
across the array). Due to the frequency drift and phase 
jitter of the oscillator, the frequency and phase of the $n$-th node in the $k$-th iteration can be $\text{written as}$  
\begin{align}\label{FreqPhase_model}
f_n(k)&=f_n(k-1)+\delta f_n\nonumber\\
\theta_n(k)&=\theta_n(k-1)+\delta\theta^f_n+\delta\theta_n,
\end{align}
in which $\delta f_n$ is the frequency drift of the oscillator at the $n$-th node which we assume is normally 
distributed as $\N\left(0,\sigma^2_f\right)$ with $\sigma_f$ representing the Allan deviation (ADEV) 
metric of the oscillator. The ADEV is modeled as $\sigma_f=f_c\sqrt{\frac{\beta_1}{T}+\beta_2 T}$ 
in which the oscillator's design parameters $\beta_1$ and $\beta_2$ are set as 
$\beta_1=\beta_2=5\times 10^{-19}$ which represent a quartz crystal oscillator 
\cite{Hassna_TWC, rashid2022frequency}. The phase error $\delta\theta^f_n$ 
in \eqref{FreqPhase_model} represents the phase offset due to the temporal variation of the frequency 
drift $\delta f_n$ between the update intervals which is $\delta\theta^f_n=-\pi T\delta f_n$ as shown in \cite{rashid2022frequency}. 
Finally, $\delta\theta_n$ in \eqref{FreqPhase_model} represents the phase 
offset due to the phase jitter of the oscillator at the $n$-th node. 
It is modeled as $\delta\theta_n\sim \N(0,\sigma_\theta)$ in which 
$\sigma_\theta=\sqrt{2\times 10^{A/10}}$ and $A$ is the integrated phase noise power of the oscillator 
which is obtained from its phase noise profile. Herein, we set 
$A=-53.46$ dB that models a typical high phase noise voltage controlled oscillator  
\cite{Serge_Access_2021, rashid2022frequency}. We assume that the 
frequency transitioning process in \eqref{FreqPhase_model} 
begins with the initial value $f_n(0)\sim\N(f_c,\sigma^2)$ in which $f_c$ is the carrier frequency and 
$\sigma=10^{-4}f_c$ denotes a crystal clock accuracy of $100$ parts per million (ppm), whereas 
the phase transitioning process starts with $\theta_n(0)\sim\mathcal{U}(0,2\pi)$ which represents the hardware 
induced initial phase offset of the oscillator.

Due to the above dynamics of the oscillators, 
the nodes need to estimate their frequencies and phases in each iteration before updating them, 
and thus the estimated 
frequency and phase of the $n$-th node in the $k$-th iteration are given by 
\begin{align}\label{FreqPhase_est}
\hat{f}_n(k)=f_n(k)+\varepsilon_f\nonumber\\
\hat{\theta}_n(k)=\theta_n(k)+\varepsilon_\theta,
\end{align}
in which the frequency estimation error $\varepsilon_f$ and the phase estimation error 
$\varepsilon_\theta$ are both normally distributed with zero mean and standard deviation $\sigma^m_f$ and 
$\sigma^m_\theta$, respectively. Since the focus of this work is on evaluating the synchronization 
performance of the algorithms, we set these standard deviations equal to their Cramer Rao lower bounds (CRLBs) 
as $\sigma^m_f=f_c\sqrt{\frac{6}{(2\pi)^2 L^3 \text{SNR}}}$ and 
$\sigma^m_\theta= \frac{2L^{-1}}{ \text{SNR}}$ \cite{Richards_radar}. 
In these equations, $L=Tf_s$ is 
the number of received signal samples collected over the time duration of length $T$ with sampling frequency $f_s$, 
and SNR represents the signal to noise ratio of the signals. 
%
\begin{figure*}[t!]
    \begin{minipage}{0.34\textwidth}
        \centering
\includegraphics[width=1.0\textwidth,height=0.65\textwidth]{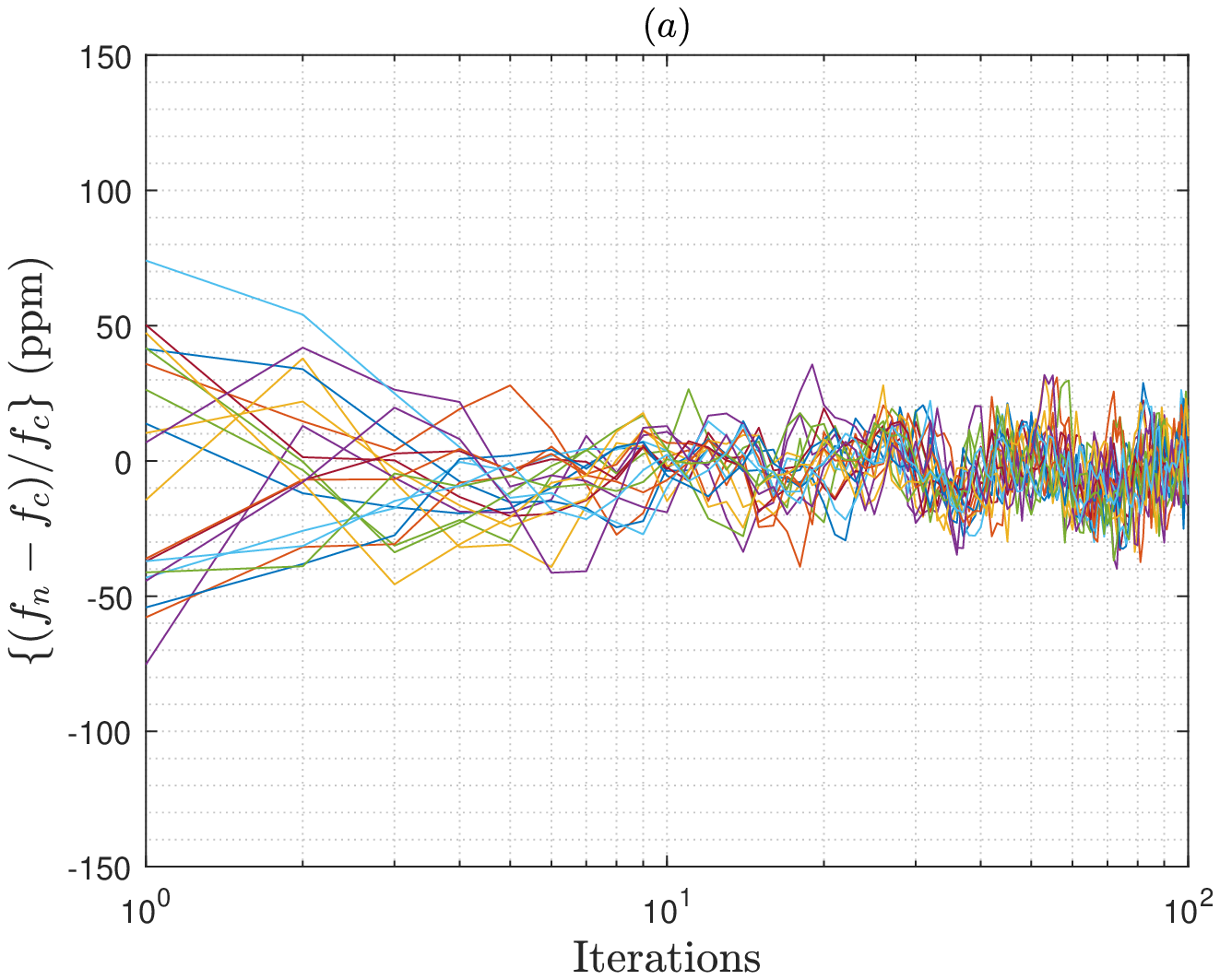}
    \end{minipage}
    \begin{minipage}{0.34\textwidth}
        \centering
\includegraphics[width=1.0\textwidth,height=0.65\textwidth]{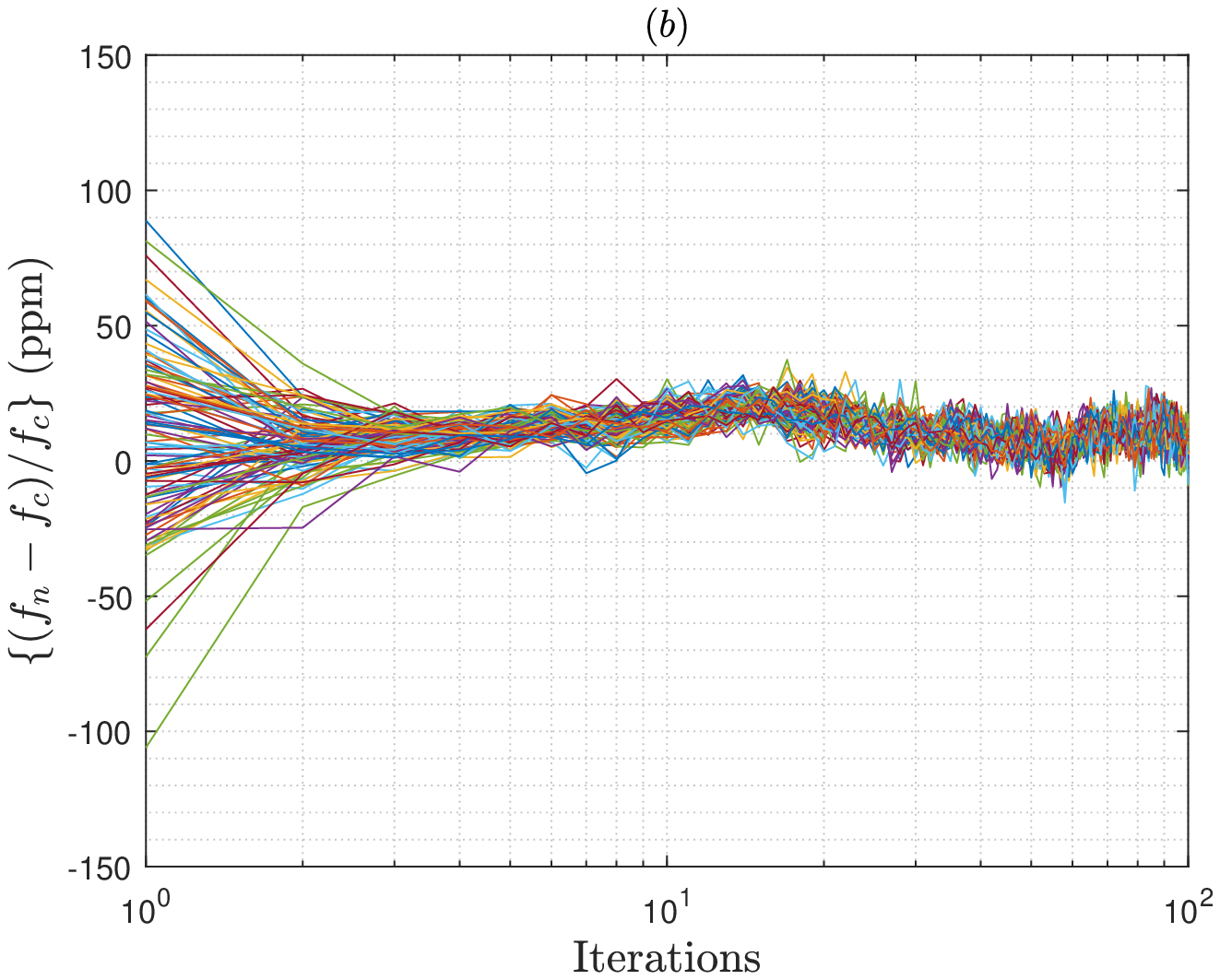}
		\end{minipage}
    \begin{minipage}{0.34\textwidth}
        \centering
\includegraphics[width=1.0\textwidth,height=0.65\textwidth]{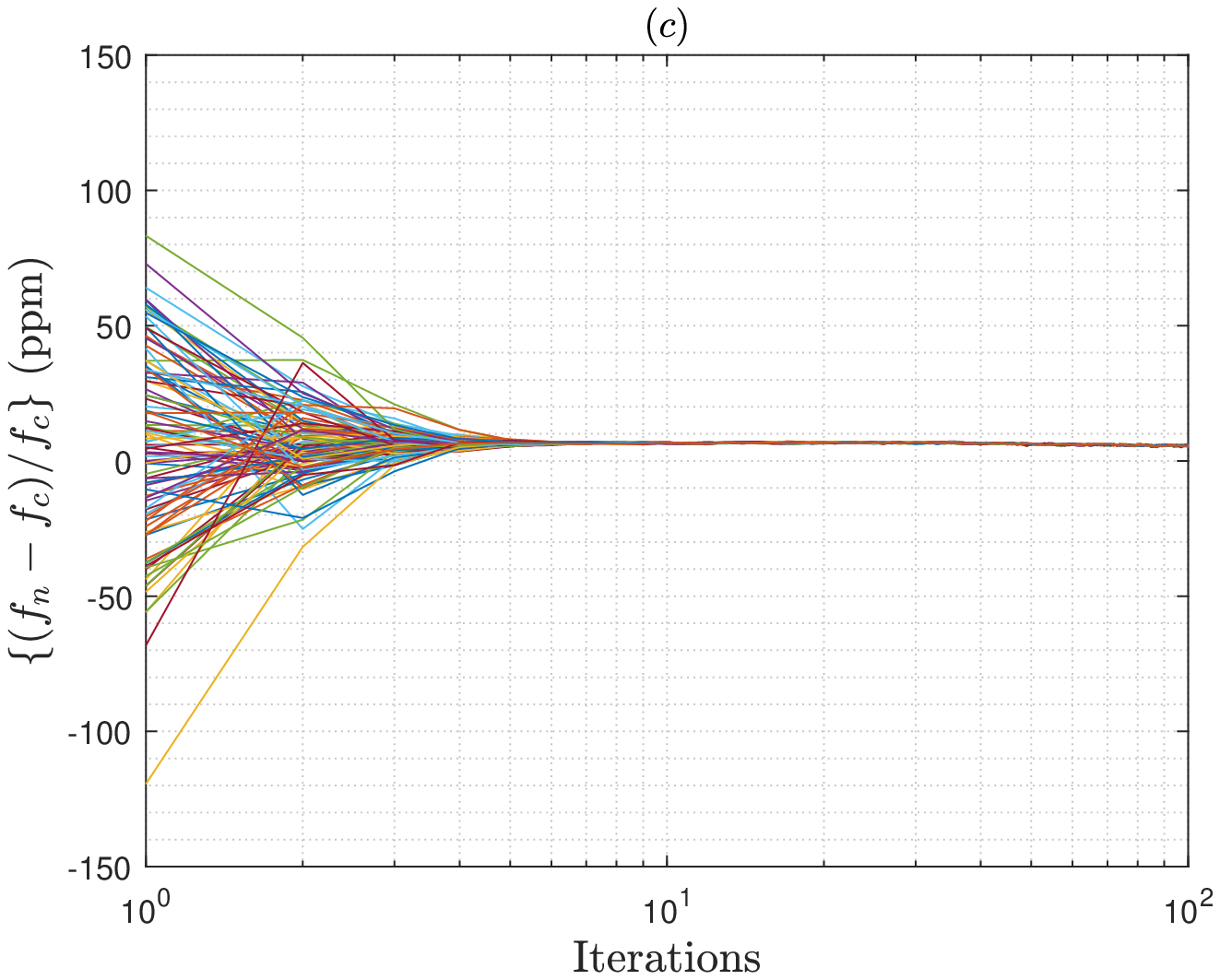}
		\end{minipage}
		\caption{Frequency errors in ppm for all the $N$ nodes in the array vs. $\text{P}_\text{s}$FPC 
		iterations for $c=0.2$ 
		and when $(a)$ $N=20$ and $\text{SNR}=0$ dB $(b)$ $N=100$ and $\text{SNR}=0$ dB, and 
		$(c)$ $N=100$ and $\text{SNR}=30$ dB.}
		\label{fig:freqs}
\end{figure*}
\begin{figure*}[t!]
    \begin{minipage}{0.34\textwidth}
        \centering
\includegraphics[width=1.0\textwidth,height=0.65\textwidth]{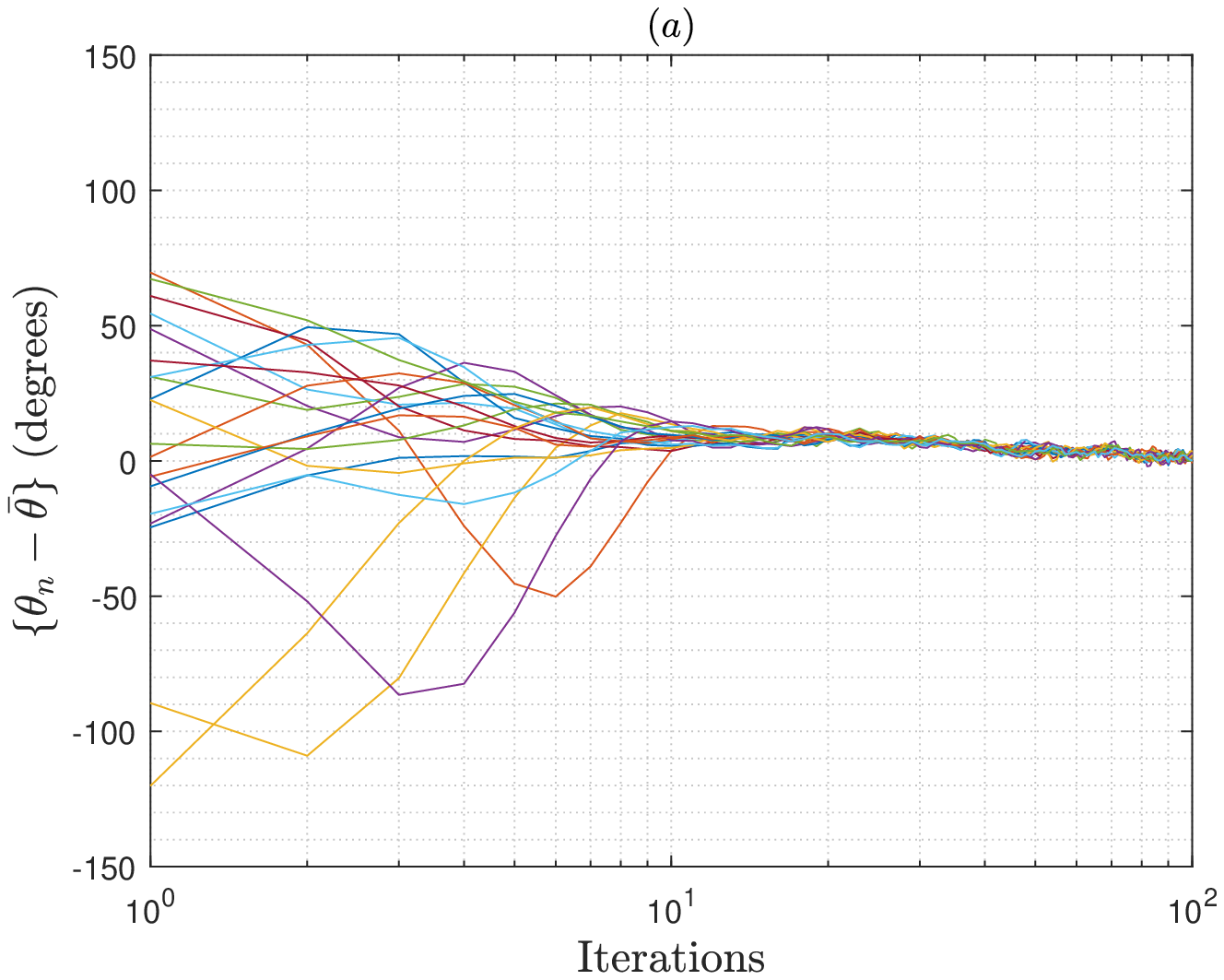}
    \end{minipage}
    \begin{minipage}{0.34\textwidth}
        \centering
\includegraphics[width=1.0\textwidth,height=0.65\textwidth]{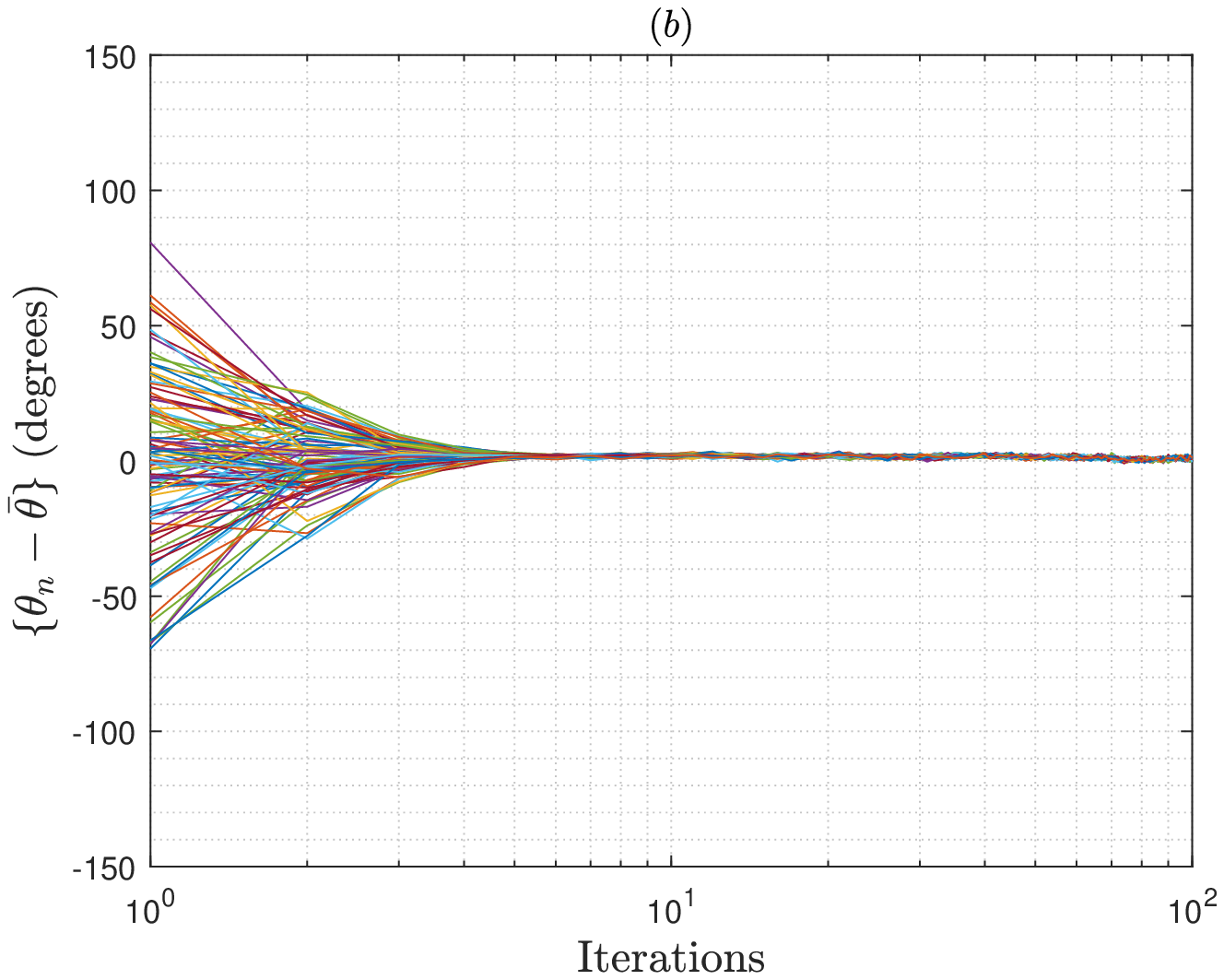}
		\end{minipage}
    \begin{minipage}{0.34\textwidth}
        \centering
\includegraphics[width=1.0\textwidth,height=0.65\textwidth]{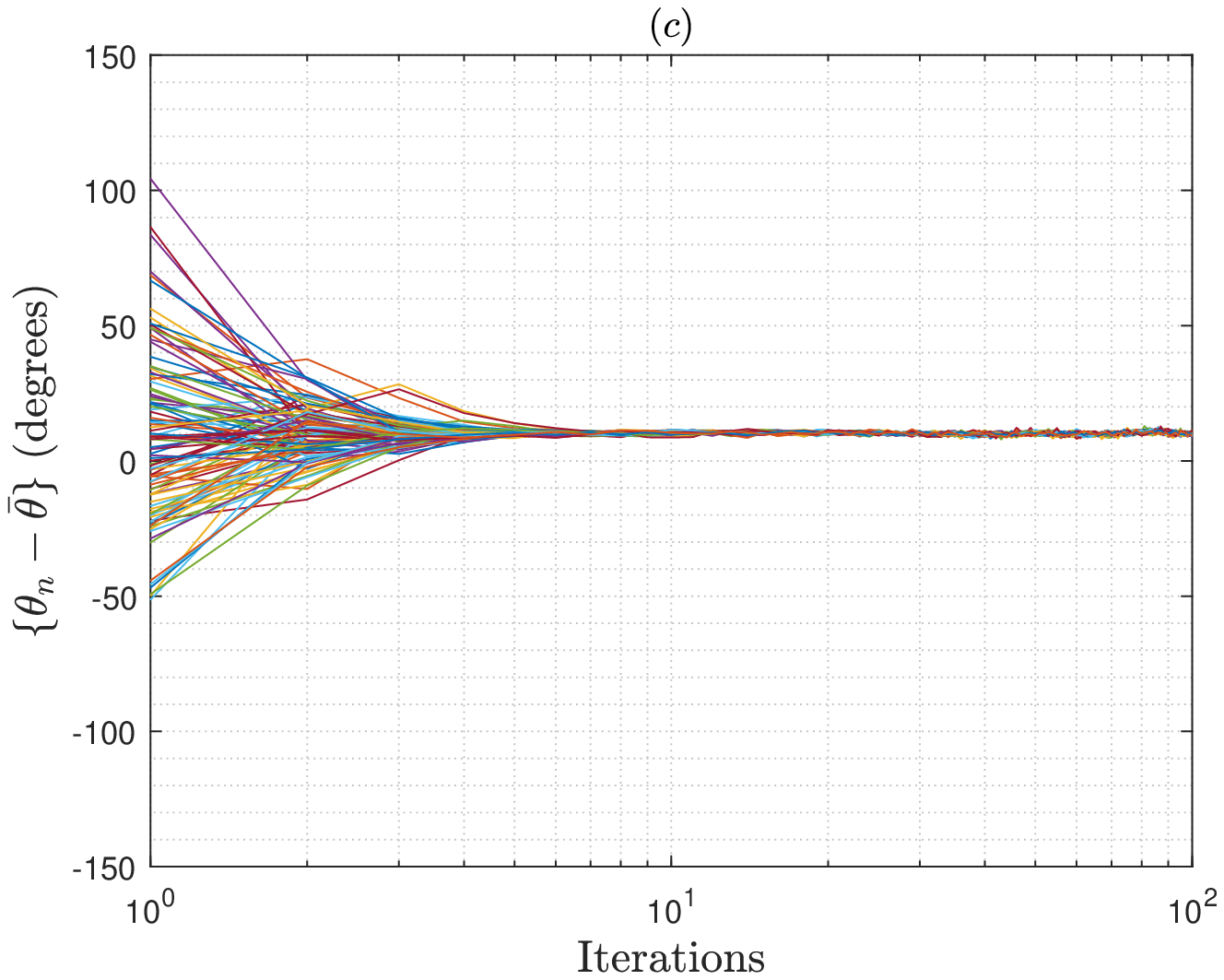}
		\end{minipage}

		\caption{Phase errors in degrees for all the $N$ nodes in the array vs. $\text{P}_\text{s}$FPC 
		iterations for $c=0.2$ 
		and when $(a)$ $N=20$ and $\text{SNR}=0$ dB $(b)$ $N=100$ and $\text{SNR}=0$ dB, and 
		$(c)$ $N=100$ and $\text{SNR}=30$ dB.}
		\label{fig:phases}
\end{figure*}

\subsection{Push-Sum Frequency and Phase Consensus Algorithm} \label{PsFPC_section}
We consider a directed array network of $N$ nodes described by a directed 
graph $\mathcal{G}=(\mathcal{V},\mathcal{E})$ as before, whereas 
the $(n,m)$-th edge in $\mathcal{E}$ 
is assigned herein a weight $w_{m,n}$. We assume that $\mathcal{G}$ 
represents a strongly connected directed graph, i.e., there exists at least one directed path from any node 
$n$ to any node $m$ in the graph with $n\neq m$. Furthermore, it is assumed that the $n$-th node in the network 
knows its in-neighbor set $\chi^{in}_n\triangleq\{v\in\mathcal{V}:w_{n,v}>0\}$ 
and its out-neighbor set $\chi^{out}_n\triangleq\{v\in\mathcal{V}:w_{v,n}>0\}$. 
The proposed algorithm is based on the push-sum consensus algorithm \cite{PS_1, Hassna_TAP}, 
and is referred to herein as the push-sum frequency and phase consensus ($\text{P}_\text{s}$FPC) algorithm. 
In the $\text{P}_\text{s}$FPC algorithm, 
the $n$-th node maintains three temporary variables 
in the $k$-th iteration, 
i.e., $x^f_n(k)$, $x^\theta_n(k)$, and $s_n(k)$, which are updated as follows. 
\begin{align}\label{temp_PsFPC}
x^f_n(k)&=\sum_{m\in\chi^{in}_n} w_{n,m} \hat{f}_m(k)s_m(k-1)\nonumber\\
x^\theta_n(k)&=\sum_{m\in\chi^{in}_n} w_{n,m} \hat{\theta}_m(k)s_m(k-1)\nonumber\\
s_n(k)&=\sum_{m\in\chi^{in}_n} w_{n,m} s_m(k-1),
\end{align}
where $s_n(0)=1$ and the weight $w_{n,m}\triangleq \frac{1}{d^{out}_m}$ if $m\in\chi^{in}_n$ and is $0$ otherwise. 
The parameter $d^{out}_m$ is the out-degree of node $m$ which 
can be found by computing the cardinality of the set $\chi^{out}_m$.  
These weights make the $N\times N$ weighting matrix $\bdW=[w_{n,m}]$ a column stochastic 
matrix that supports an average consensus \cite{PS_2}. The parameters $\hat{f}_m(k)$ and $\hat{\theta}_m(k)$ represent 
the estimated frequency and phase of the $m$-th node in the $k$-th iteration, 
and they are multiplied by the weighting factor $s_m(k-1)$ from the previous iteration. This multiplication is required 
to recover the temporary variables $x^f_n(k-1)$ and $x^\theta_n(k-1)$ 
taking into account the fact that at the beginning of each iteration, the nodes observe only the estimated frequencies and 
phases after their updated values in the previous iteration are influenced by the oscillator drifts and the phase jitters. 
The updated frequency and phase of the $n$-th node 
in the $k$-th iteration are computed by
\begin{align}\label{upd_PsFPC}
f_n(k)&=\frac{x^f_n(k)}{s_n(k)}\nonumber\\
\theta_n(k)&=\frac{x^\theta_n(k)}{s_n(k)},
\end{align} 
The above steps are repeated iteratively at each node until the algorithm converges. The pseudo-code of the 
$\text{P}_\text{s}$FPC algorithm is provided in Algorithm \ref{algo_1}.
\begin{algorithm}[t!]\label{algo_1}
  \footnotesize
\DontPrintSemicolon
\SetKwInput{KwPara}{Input}
\KwPara{$k=0$, define $s_n(0)=1$ for all $n=1,2,\ldots,N$.} 
\While{convergence criterion is not met} 
{
	$k=k+1$\\
For each node $n$:
\begin{enumerate}
\item[a)] Use the observed $\hat{f}_m(k)$ and $\hat{\theta}_m(k)$ for all $m\in\chi^{in}_n$ 
and \\update  
$x^f_n(k)$, $x^\theta_n(k)$, and $s_n(k)$ using \eqref{temp_PsFPC}.
\item[b)] Update $f_n(k)$ and $\theta_n(k)$ using \eqref{upd_PsFPC}. 
\end{enumerate}

}
\KwOut{$f_n(k)$ and $\theta_n(k)$ for all $n=1,2,\ldots,N$}
\caption{$\text{P}_\text{s}$FPC Algorithm}
\end{algorithm}
\subsection{Residual Phase Error Analysis}
Herein, we theoretically analyze the residual phase errors of the nodes upon the convergence of the 
$\text{P}_\text{s}$FPC algorithm. To this end, let at the $I$-th iteration, the estimated frequencies and 
phases of the nodes be written in the short-hand vector form as $\hat{\bdz}(I)=\bdz(I-1)+\bde_I$ in which 
$\bdz\in\{\bdf,\bdtheta\}$ and $\bde_I\sim\N\left(\bdzero,\sigma^2_e\bdI_N\right)$. Note that by using 
\eqref{FreqPhase_model} and \eqref{FreqPhase_est}, it can be seen that 
when $\bdz=\bdf$ then the error variance is $\sigma^2_e=\sigma^2_f+\left(\sigma^m_f\right)^2$, and when 
$\bdz=\bdtheta$ then the variance is $\sigma^2_e=\pi^2 T^2\sigma^2_f+\left(\sigma^m_\theta\right)^2+\sigma^2_\theta$.   
Thus, at the $I$-th iteration, the temporary variables 
in \eqref{temp_PsFPC} can be alternatively written as
\begin{align}\label{res_1}
x^z_n(I)&=\left[\bdW\left(\hat{\bdz}(I)\odot \bds(I-1)\right)\right]_n\nonumber\\
s_n(I)&=\left[\bdW\bds(I-1)\right]_n,
\end{align}
where $\odot$ is the Hadamard product between the two vectors, 
and the operation $[.]_n$ selects the $n$-th element of the resulting vector. Inserting $\hat{\bdz}(I)$ in 
\eqref{res_1} and using backward recursion, it can be reduced as follows
\begin{align}\label{res_2}
x^z_n(I)=\left[\bdW\left({\bdz}(I-1)\odot \bds(I-1)\right)+\bdW\left(\bde_I\odot\bds(I-1)\right)\right]_n\nonumber\\
=\left[\bdW^I\left({\bdz}(0)\odot \bds(0)\right)+\sum^I_{i=1}\bdW^i\left(\bde_{I-(i-1)}\odot\bds(I-i)\right)\right]_n\nonumber\\
\end{align}
Similarly we have $s_n(I)=\left[\bdW^I\bds(0)\right]_n$. For a column stochastic 
matrix $\bdW$, it is shown in \cite{Wang_2021} that as $I\ra\infty$, $\bdW^I=\bdpi \bdone^T$ where 
$\bdpi>0$ is the stationary distribution of the 
random process with transition matrix $\bdW$ that satisfies $\bdW\bdpi=\bdpi$ with $\bdone^T\bdpi=1$. 
Thus from \eqref{upd_PsFPC}, the 
updated frequency or phase of node $n$ at the $I$-th iteration can be written in the shorthand form as 
\begin{align}\label{res_3}
&z_n(I)=\nonumber\\
&\frac{1}{\pi_n N}\left[\bdW^I\left({\bdz}(0)\odot \bds(0)\right)+\sum^I_{i=1}\bdW^i\left(\bde_{I-(i-1)}\odot\bds(I-i)\right)\right]_n,
\end{align}
where to get \eqref{res_3} we set $\bds(0)=\bdone$ (as in Section \ref{PsFPC_section}) that gives 
for a large $I$, $s_n(I)=\bdW^I\bds(0)=\pi_n N$ in which $\pi_n$ is the $n$-th element 
of $\bdpi$. Similarly, for larger $I$, the first summand in \eqref{res_3} 
reduces to $\frac{1}{N}\sum^N_{n=1}z_n(0)$, i.e., it gives the average of the initial frequencies and phases of 
the nodes, whereas the second summand in \eqref{res_3} represents the residual total phase error due to the frequency and 
phase offset errors induced at the nodes in every iteration. To quantify this residual phase error for node $n$ 
after $I$ iterations, we find the variance of the following term
\begin{align}
&\frac{1}{\pi_n N}\left[\sum^I_{i=1}\left(\bdW^i-\bdpi\bdone^T\right)\left(\bde_{I-(i-1)}\odot\bds(I-i)\right)\right]_n\nonumber\\
&=\frac{1}{\pi_n N}\left[\sum^I_{i=1}\left(\bdW-\bdpi\bdone^T\right)^i\left(\bde_{I-(i-1)}\odot\bds(I-i)\right)\right]_n\end{align}
where to get the second equality in the above equation, we used the following two facts. Firstly, a column stochastic 
$\bdW$ implies that $\bdone^T\bdW^i=\bdone$, and secondly, we have 
$\left(\bdI_N-\bdpi\bdone^T\right)^i=\left(\bdI_N-\bdpi\bdone^T\right)$ as it is a projection matrix. 
Now let $\lambda_2$ be the modulus of the second largest eigenvalue of the mixing matrix $\bdW$. 
Then it can be conveniently shown in a few steps that 
$\lambda^i_2 \bdI_N-\left(\bdW-\bdpi\bdone^T\right)^i$ is a positive semi-definite matrix. Thus, 
using this identity along with assuming that the frequency and phase offset errors induced 
at the nodes in every iteration are mutually independent across the array, 
the variance of the residual phase error for node $n$ can be solved as  
\begin{align}\label{res_err_1}
\sigma^2_{n,\text{residual}} &=\frac{\sigma^2_e}{|\pi_n N|^2}\sum^I_{i=1}\lambda^{2i}_2|s_n(I-i)|^2,
\end{align}
Since for large $I$, the scalars $s_n(I)\rightarrow \pi_n N$, it implies that 
we can write the phase error in \eqref{res_err_1} as 
$\sigma^2_{n,\text{residual}}= \sigma^2_e \sum^{I^{'}}_{i=1}\lambda^{2i}_2+C$ where 
the constant $C$ is close to zero for the larger arrays. Consequently, for sparsely connected 
arrays, $\lambda_2$ is close to $1$ and $\sum^I_{i=1}\lambda^{2i}_2\gg 1$ which results in a higher residual 
phase error, but as the connectivity increases, $\lambda_2\ra 0$ and the residual phase error decreases. 
This variation in the total phase error as a function of the change in the array connectivity is illustrated through 
simulations in Section \ref{KF_sims}.   

\subsection{Simulation Results}\label{sim_results_PsFPC}
For validation purposes, we examine the frequency and phase synchronization performances 
of the proposed $\text{P}_\text{s}$FPC algorithm through simulations. 
To this end, we consider an array network of $N$ nodes that is randomly generated with connectivity $c$ 
between the nodes. The parameter 
$c$ is defined as the total number of existing connection in the generated network divided by the total 
number of all possible connections $\frac{N}{2}(N-1)$. 
Thus, $c\in [0,1]$ where a smaller value of $c$ implies a sparsely connected array network with a few 
connections per node, and a larger value of $c$ defines a highly connected array network with many connections per node. The 
average number of connections per node in a network is given by $D=c(N-1)$. The carrier frequency of the nodes 
is chosen in this analysis to be $f_c=1$ GHz, and the sampling frequency is set to $f_s=10$ MHz throughout this paper. As the update interval 
for distributed phased arrays is usually on the order of ms, we set $T=0.1$ ms for all the simulated cases in this paper. 

Figs. \ref{fig:freqs} and \ref{fig:phases} show the frequencies and phases for all the $N$ nodes in the array 
relative to their average values vs. the number of iteration of 
the $\text{P}_\text{s}$FPC algorithm from a single trial, when 
a moderately connected array network with connectivity $c=0.2$ is generated for the different number of nodes $N$ in 
the array. The SNR of the observed signals is either set as $0$ dB or $30$ dB. These figures show that with the 
increase in the number of 
iterations, both frequencies and phases of the nodes converge to the average of their initial 
values for all the considered cases. 
The residual error upon convergence results from the oscillators' frequency drifts and phase jitters, as 
well as the frequency and phase estimation errors induced at the nodes. 
Thus, when the SNR increases to $30$ dB for $N=100$ nodes, it is observed that the residual error also 
decreases significantly upon the convergence due to the decrease in the estimation errors. 
\begin{figure*}[tp]
    \begin{minipage}{0.48\textwidth}
        \centering
\includegraphics[width=0.99\textwidth]{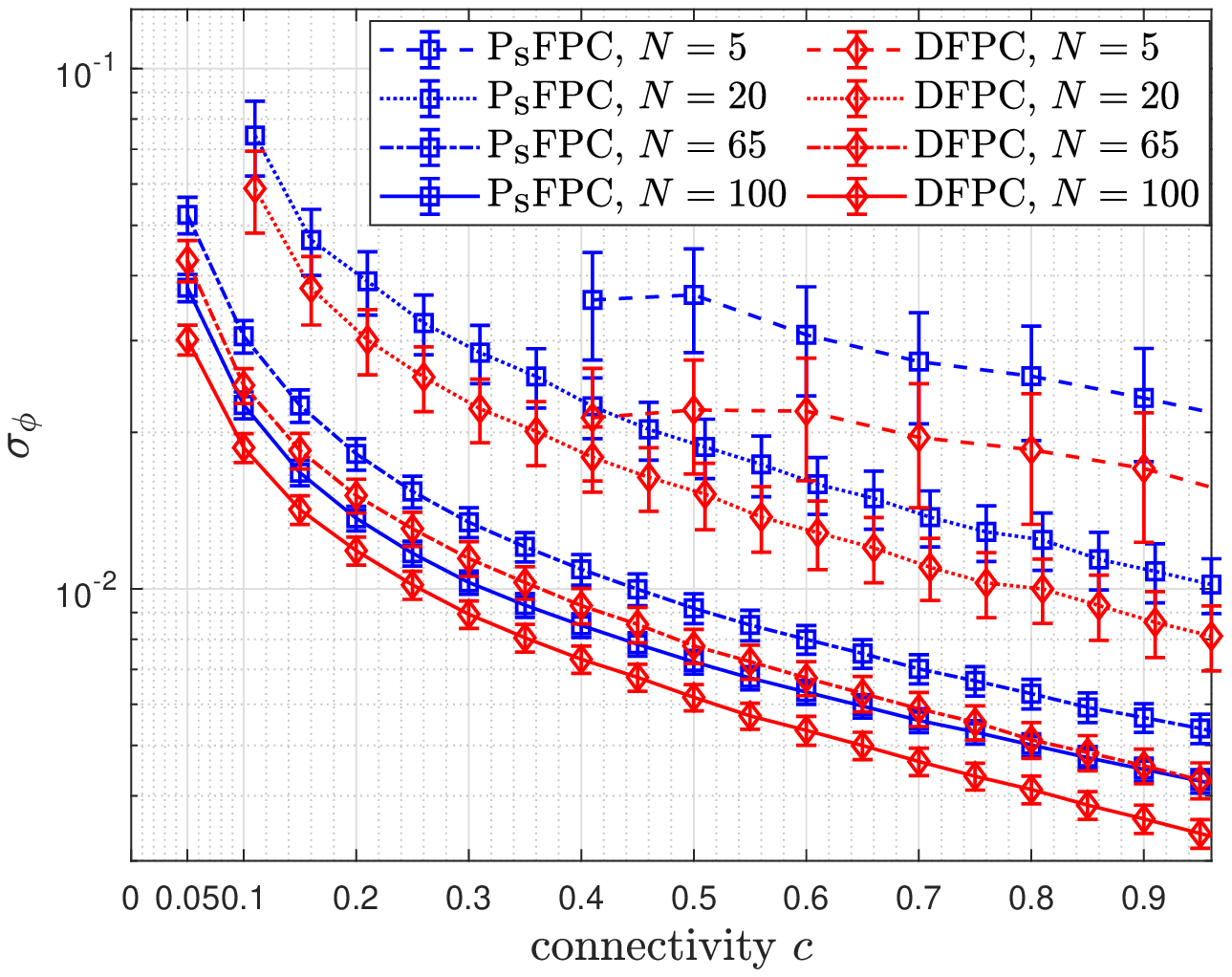}
	\caption{Standard deviation of the total phase errors of the $\text{P}_\text{s}$FPC and DFPC algorithms  vs. 
	the connectivity $c$ in the array for different $N$ values when SNR$=30$ dB is assumed.}
	\label{fig:PsFPC_stderr_vs_c}
		\end{minipage}\hspace{.025\linewidth}
    \begin{minipage}{0.48\textwidth}
        \centering
\includegraphics[width=0.99\textwidth]{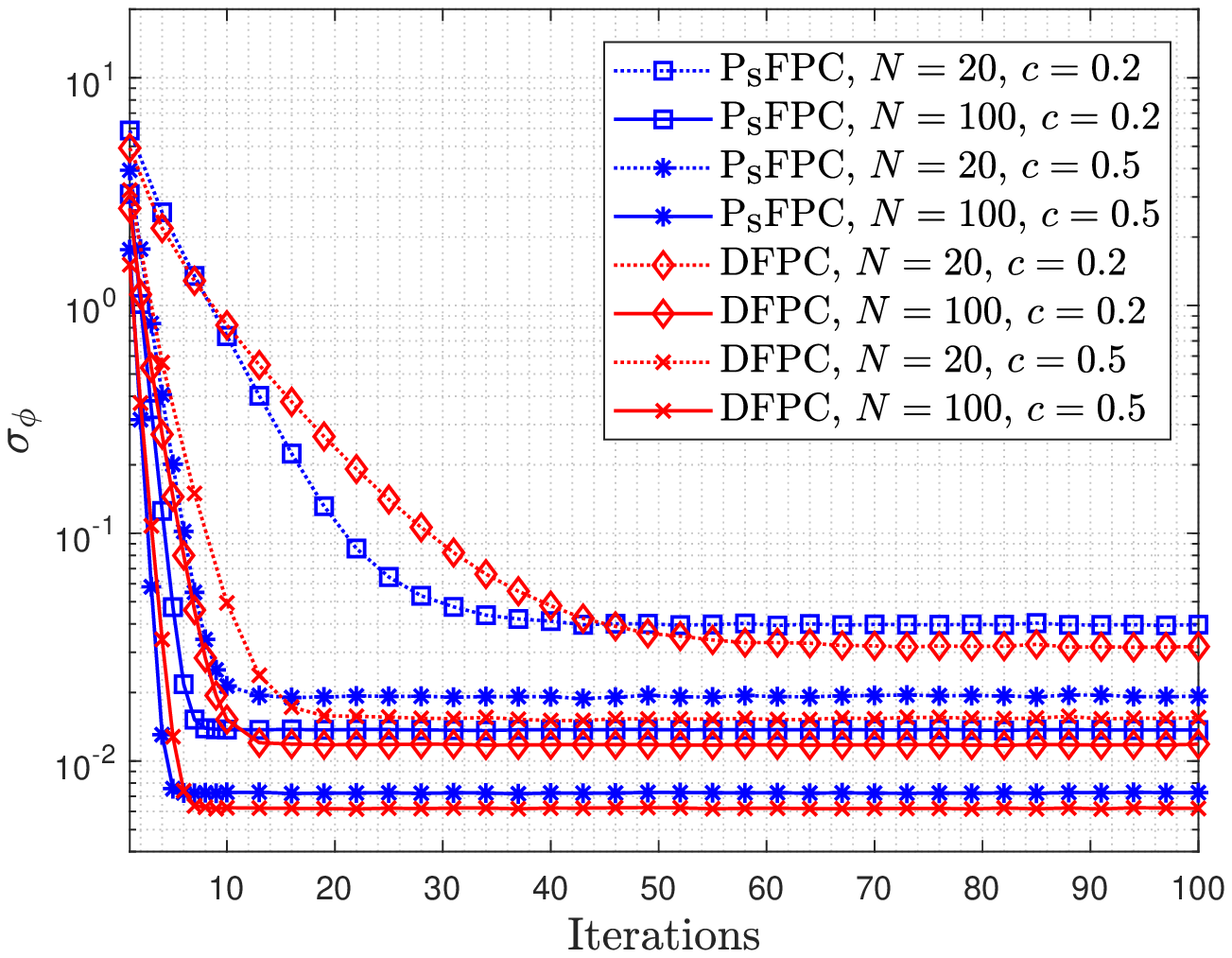}
\caption{Standard deviation of the total phase errors of the $\text{P}_\text{s}$FPC and DFPC algorithms  vs. 
	the number of iterations for different $N$ and $c$ values when SNR$=30$ dB is assumed.}	
	\label{fig:PsFPC_stderr_vs_iter}
		\end{minipage}\hspace{.01\linewidth}
\end{figure*}

Now let the total residual phase error of node $n$ be denoted by $\delta \phi_n=2\pi\delta f_n T+2\pi\varepsilon_f T+\delta\theta^f_n+\delta\theta_n+\varepsilon_\theta$. Thus, 
in Fig. \ref{fig:PsFPC_stderr_vs_c}, we 
plot the standard deviation of the total phase errors $\delta \phi_n$
for all the $N$ nodes in the array for the $\text{P}_\text{s}$FPC algorithm 
by varying the connectivity $c$ between the nodes and the number of nodes $N$ in the array, and when SNR$=30$ dB is assumed. 
For the comparison, we generate the bidirectional array networks for each $c$ and $N$ values and 
show the standard deviation of the total phase error of the DFPC algorithm 
proposed in \cite{rashid2022frequency} for such array networks. Note that the standard deviation of the DFPC-based algorithms in 
this work provides the 
lower bounds on the standard deviation of the total 
phase error of the $\text{P}_\text{s}$FPC-based algorithms, and the slight increase in 
the phase error of the $\text{P}_\text{s}$FPC-based algorithms as compared to the DFPC-based algorithms 
is due to the decrease in the number of connection per node in the directed networks which effects the accuracy of the local 
averages computed per node throughout the network as expected. 
For this figure, both $\text{P}_\text{s}$FPC and DFPC algorithms were run for $100$ iterations and 
the final standard deviations of the total phase errors were averaged over $10^3$ trials
The 
center of each point in the error bar plot is the average value of the samples and the 
length of the bar defines the standard deviation around the average value. 
It is observed that for each $N$ value, as the connectivity $c$ between the nodes 
increases then the total phase error of the $\text{P}_\text{s}$FPC and DFPC algorithms decreases proportionately.  Furthermore, 
an increase in the number of nodes $N$ in the array, for a given $c$ value, 
also decreases the total phase errors of both algorithms. 
As discussed in the previous subsection, the decrease in the residual phase error of $\text{P}_\text{s}$FPC 
with the increase in the $c$ or $N$ values is because the network 
algebraic connectivity $\lambda_2$ decreases 
which in turn reduces the residual phase of $\text{P}_\text{s}$FPC. 
The residual phase error of DFPC is also a function of $\lambda_2$ as shown 
in \cite{rashid2022frequency} and thus the same trend is observed for 
the DFPC algorithm as well. 

Finally, in Fig. \ref{fig:PsFPC_stderr_vs_iter}, we compare the convergence speeds of the $\text{P}_\text{s}$FPC and 
DFPC algorithms by plotting the standard deviation of the total phase error vs. the number of iteration of the algorithms 
when the two different $N$ and $c$ values are assumed and the SNR is set to $30$ dB. 
It is observed that either when $N$ increases from $20$ to $100$ for a given $c$ value, or when 
$c$ increases from $0.2$ to $0.5$ for a given $N$ value, the $\text{P}_\text{s}$FPC algorithm converges faster. Specifically, 
for $N=20$ and $c=0.2$, $\text{P}_\text{s}$FPC converges in $32$ iterations, but for this $N$ value, 
when $c$ increases to $0.5$, $\text{P}_\text{s}$FPC converges in about $11$ iterations. 
Moreover, for $c=0.5$ and $N=100$ nodes, $\text{P}_\text{s}$FPC 
converges in about $5$ iterations. This is because as either $c$ or $N$ increases then the average number of 
connections per node in the network, i.e., the $D$ value, increases which in turn quickly stabilizes the local averages computed 
at the nodes. Although the same trend is also observed for the DFPC algorithm in all 
the considered cases in this figure, particularly it is observed that 
the $\text{P}_\text{s}$FPC algorithm converges faster than the 
DFPC algorithm in all cases. 
This is because the network is bidirectional for the DFPC algorithm, and as such there are 
more connections per node in the network which in turn delays the convergence of the local averages computed at the nodes. 
{Note that the $\text{P}_\text{s}$FPC algorithm can be deployed for a bidirectional network as well, 
in which case it results in 
the same synchronization and convergence performance as that of the DFPC algorithm}; however, 
the DFPC algorithm does not converge in case of directed networks due to its dependence on 
the average consensus algorithm \cite{Fast_MC_2004} as discussed before. In practice, since the failure of a 
link may not be easily detectable, this means that 
the $\text{P}_\text{s}$FPC algorithm is a preferred choice over the DFPC algorithm. 
\section{Mitigating Residual Phase Errors with Online Expectation-Maximization and Kalman Filtering}\label{EM_section}

In order to reduce the residual phase error of the $\text{P}_\text{s}$FPC algorithm, 
we propose to use Kalman filtering (KF) at the nodes that computes the 
MMSE estimates of the frequencies and phases 
in each iteration to mitigate the offset errors. KF is applicable in scenarios where the 
state transitioning model follows a first-order Markov process, 
and the observations are a linear function of the unobserved state \cite{Sarkka_2013}.
KF has been used for the time synchronization between the nodes in \cite{KF-TS_2015, KF-TS_2019}, and 
for the frequency and phase synchronizations between the nodes in a distributed antenna arrays 
in \cite{rashid2022frequency}. However, an implicit 
assumption in using KF with the synchronization algorithms 
in \cite{KF-TS_2015, KF-TS_2019, rashid2022frequency} is that 
the innovation noise and the 
measurement noise covariance matrices are known to the Kalman filter. 
As discussed in Section \ref{intro_section}, 
these matrices are usually unknown and must be estimated for the KF algorithm. Thus, in this section, 
we propose to use an online expectation-maximization (EM) algorithm \cite{Cappe_EM_2009} that iteratively computes 
the maximum likelihood estimates of these noise matrices for Kalman filtering. 
The EM algorithm is integrated with the KF and $\text{P}_\text{s}$FPC algorithms, and the resulting algorithm 
is referred to herein as EM-KF-$\text{P}_\text{s}$FPC. 
Note that our proposed EM algorithm iteratively computes the ML estimates 
in an online fashion wherein the estimates are updated instantaneously as the new measurements are recorded. On the other 
hand, the EM algorithm of \cite{Batch_EM_2022} assumes a batch mode technique in which a batch of measurements are 
recorded a priori, and then the EM algorithm is run iteratively on that set of measurements until its convergence. 
This batch-mode EM algorithm is not applicable in our considered synchronization problem wherein we want to update 
the nodes in an online manner to avoid larger oscillators drifts and to simultaneously 
perform the coherent operation during the synchronized interval. Similarly, a variational Bayes (VB) method based 
hybrid consensus on measurement and consensus on information (VB-HCMCI) algorithm was also proposed in \cite{VB_HCMCI_2021}, 
wherein the unknown noise 
matrices are estimated in an online manner using VB; however, this VB-based algorithm additionally requires 
a consensus on measurements and a consensus on the predicted information between the nodes at each time instant 
to result in an improved 
performance, and they also need to perform 
multiple iterations at each time instant for the VB's convergence which makes them computationally more expensive than 
EM-KF-$\text{P}_\text{s}$FPC. We note that performing multiple iterations at each time instant increases the update 
interval of the nodes in a practical distributed phased array system which is not favorable to avoid the decoherence between the nodes due 
to the larger oscillator drifts. 
Furthermore, these VB-based algorithms are developed for the undirected networks using the average consensus algorithm 
\cite{Fast_MC_2004}, and thus do not converge for the directed ones. 

Now, to develop the EM and KF algorithms, we start with 
describing the temporal variation of the frequencies and phases of the nodes via a state-space model as follows.

\subsection{State-Space Model}

To write the state-space equation for the $n$-th node, let at the $k$-th time instant its 
state vector be given by $\bdx_n(k)=[f_n(k),\theta_n(k)]^T$ in which 
$f_n(k)$ and $\theta_n(k)$ represent its frequency and phase, respectively. 
Using the frequency drift and phase jitter models as described in Section \ref{offset_modeling}, 
the state-space equation for the $n$-th node is written as 
\begin{equation}\label{state_eqn}
\bdx_n(k)=\bdx_n(k-1)+\bdu_n,
\end{equation}
in which the innovation noise vector is defined as 
$\bdu_n=[\delta f_n,\delta\theta^f_n+\delta\theta_n]^T$, and 
we assume that $\bdu_n$ is normally distributed with zero mean and the correlation matrix $\bdQ$ which 
is given by 
\begin{equation}\label{Q_mtx_eqn}
\bdQ=\Exp[\bdu_n\bdu^T_n]=
\left[
\begin{matrix}
\sigma^2_f && -\pi T\sigma^2_f\\
-\pi T\sigma^2_f && \pi^2T^2\sigma^2_f+\sigma^2_\theta
\end{matrix}
\right],
\end{equation}

Next to write the observation equation, let the frequency and phase estimates of the signal for the $n$-th node 
be written in the vector form as $\bdy_n(k)=\left[\hat{f}_n(k),\hat{\theta}_n(k)\right]^T$,  
then in terms of the estimation errors, this observation vector is defined as 
\begin{equation}\label{obs_eqn}
\bdy_n(k)=\bdx_n(k)+\bdv_n,
\end{equation}
in which the measurement noise vector $\bdv_n$ is given by 
$\bdv_n=\left[\varepsilon_f,\varepsilon_\theta\right]^T$ in which $\varepsilon_f$ and 
$\varepsilon_\theta$ denote the frequency and phase estimation errors, respectively. 
We assume that $\bdv_n$ is also normally distributed with zero mean and the correlation matrix 
$\bdSigma$ which is given by
\begin{equation}\label{Sigma_mtx_eqn}
\bdSigma=\Exp[\bdv_n\bdv^T_n]=\left[
\begin{matrix}
\left(\sigma^m_f\right)^2 && 0\\
0 && \left(\sigma^m_\theta\right)^2
\end{matrix}
\right],
\end{equation} 
in which the standard deviations $\sigma^m_f$ and $\sigma^m_\theta$ define the marginal distributions on the 
frequency and phase estimation errors, respectively. 

In the above state-space model in \eqref{state_eqn} and \eqref{obs_eqn},
the measurement noise vector $\bdv_n$ is independent of the state vector $\bdx_n(k)$ and the 
innovation noise vector $\bdu_n$, thus the state vector estimate at each time instant can be easily obtained 
by using Kalman filtering as described below. 

\subsection{Kalman Filtering}\label{KF_section}

Kalman filtering is an online estimation algorithm that estimates the state vector $\bdx_n(k)$ of 
the $n$-th node at the current time instant $k$ 
by using all the observations up to the present time. This process involves sequentially computing the 
prediction-update step and the time-update step at each time instant. 
To this end, let the state vector estimate of the $n$-th node 
at time instant $k-1$ be given by the vector $\bdm_{n,k-1}(k-1)$ whereas its error covariance matrix is 
defined by the matrix $\bdV_{n,k-1}(k-1)$. As annotated in the subscripts, 
these parameters in KF are computed by using the observations up to time $k-1$.  
Thus, in the prediction-update step at time instant $k$, we use the linear transformation 
in \eqref{state_eqn} and predict the new state vector estimate and the corresponding error 
covariance matrix as follows
\begin{align}\label{pred_upd}
\bdm_{n,k-1}(k)&=\bdm_{n,k-1}(k-1)\nonumber\\
\bdV_{n,k-1}(k)&=\bdV_{n,k-1}(k-1)+\bdQ,
\end{align}

In the time-update step, an a priori normal distribution is assumed on the state vector $\bdx_n(k)$ where 
its mean and covariance are set equal to the 
predicted estimate vector and the error covariance matrix in \eqref{pred_upd}, respectively.  
This a priori distribution is then used to compute the a posteriori mean and covariance of the state vector 
given the current observation $\bdy_n(k)$ by using 
\begin{align}
\bdm_{n,k}(k)
&=\bdm_{n,k-1}(k)+\bdK_n(k)\left(\bdy_n(k)-\bdm_{n,k-1}(k)\right)\nonumber\\
\bdV_{n,k}(k)&=\bdV_{n,k-1}(k)-\bdK_n(k)\bdV_{n,k-1}(k),\label{time_upd_mV}
\end{align}
where the Kalman gain matrix $\bdK_n(k)$ is given by
\begin{equation}\label{KF_gain}
\bdK_n(k)=\bdV_{n,k-1}(k)\left(\bdV_{n,k-1}(k)+\bdSigma\right)^{-1},
\end{equation}

The mean vector $\bdm_{n,k}(k)$ in \eqref{time_upd_mV} gives the MMSE estimate of the state vector at 
time instant $k$ and the matrix $\bdV_{n,k}(k)$ defines its error covariance matrix. These means and covariances 
of the state vectors are used in the EM algorithm derived below to compute the maximization step in every iteration.

\subsection{Online Expectation-Maximization Algorithm}\label{EM_sectn}

The prediction update and time update steps of KF in \eqref{pred_upd} and \eqref{time_upd_mV} 
assume that the innovation noise covariance matrix $\bdQ$ and the measurement noise covariance matrix 
$\bdSigma$ are known a priori. As discussed earlier, these model parameters are unknown in 
practice and must be estimated for Kalman filtering. Let $\bdTheta=\{\bdQ,\bdSigma\}$ 
denote the set of these model parameters, then given all the instantaneous observations of the node $n$ 
up to the present time $K$ as denoted by $\bdy^{1:K}_n$, the marginal log-likelihood of $\bdTheta$ is written 
as 
\begin{align}\label{ll_eqn}
&\ell(\bdTheta)=\ln \sum_{\bdx^{1:K}_n} p\left(\bdy^{1:K}_n,\bdx^{1:K}_n;\bdTheta\right)\nonumber\\
&=\ln \sum_{\bdx^{1:K}_n} \prod^K_{k=1}\left[p\left(\bdy_n(k)|\bdx_n(k);\bdSigma\right)p\left(\bdx_n(k)|\bdx_n(k-1);\bdQ\right)\right],
\end{align} 
where the conditional distributions in \eqref{ll_eqn} can be easily found from \eqref{state_eqn} and 
\eqref{obs_eqn} by shifting the distributions of the noise vectors to the given state vectors. 
Note that due to the normally distributed conditional 
distributions and 
due to the log of the summation in \eqref{ll_eqn}, the above objective function is a non-concave function of $\bdTheta$, 
and directly maximizing it 
with respect to $\bdTheta$ does not provide a closed-form solution for estimating $\bdQ$ and $\bdSigma$. 
An expectation-maximization (EM) algorithm \cite{Dempster_EM, Cappe_EM_2009} is an iterative algorithm that 
often results in the closed-form update equations for the unknown parameters and maximizes the marginal 
log-likelihood function in every iteration 
until convergence to its local maximum or saddle point \cite{Bishop}. 
An EM algorithm starts with an initial estimate of the unknown parameters, thus if $\bdTheta^{(l-1)}$ is the 
estimate at its $(l-1)$-st iteration, in the $l$-th iteration it computes 
an expectation step (E-step) and a maximization step (M-step). 
In the E-step, it computes the expectation of the complete data log-likelihood function using the 
estimate of $\bdTheta$ from the previous iteration as follows.
\begin{align}\label{Batch_Estep}
L_n\left(\bdTheta;\bdTheta^{(l-1)}\right)=
\Exp\left[\ln p\left(\bdy^{1:K}_n, \bdx^{1:K}_n;\bdTheta\right)\big| \bdy^{1:K}_n,\bdTheta^{(l-1)}\right],
\end{align}
where the expectation in \eqref{Batch_Estep} is a conditional expectation given the observations $\bdy^{1:K}_n$ 
and the estimate $\bdTheta^{(l-1)}$. 
In the M-step, it maximizes this objective function with respect to $\bdTheta$ to 
get its new estimate by solving 
\begin{align}\label{Mstep_eqn}
\bdTheta^{(l)}=\argmax_{\bdTheta} L_n \left(\bdTheta;\bdTheta^{(l-1)}\right),
\end{align}
The above E-step and M-step are repeated iteratively until the convergence is achieved. 
The EM algorithm in \eqref{Batch_Estep} and \eqref{Mstep_eqn} describes a batch-mode EM algorithm that is 
proposed in \cite{Dempster_EM, Batch_EM_2022} where 
first the entire data set up to the time instant $K$ is 
collected, and then the algorithm is run iteratively 
on this data set until convergence is achieved. 
This batch-mode EM algorithm is applicable in cases where the cost of collecting the entire data set at once is 
affordable, whereas in our synchronization task due to the oscillators instantaneous frequency drifts 
and phase jitters, we aim to instantaneously estimate $\bdTheta$ and update 
the frequencies and phases of the nodes in every iteration to synchronize these parameters 
across $\text{the array}$. 

To instantaneously estimate $\bdTheta$ from the observed data, an 
online version of the EM algorithm is proposed in \cite{Cappe_EM_2009, Schwartz_2015} where  
the E-step in \eqref{Batch_Estep} is replaced by a stochastic approximation step 
in the $k$-th iteration as follows.
\begin{align}\label{online_Estep_eqn}
L_{n,k}&\left(\bdTheta;\bdTheta^{(k-1)}\right)=L_{n,k-1}\left(\bdTheta;\bdTheta^{(k-1)}\right)+
\gamma_k\left\{\ell_{n,k}\left(\bdTheta;\bdTheta^{(k-1)}\right)-
L_{n,k-1}\left(\bdTheta;\bdTheta^{(k-1)}\right)\right\},
\end{align}
where $\gamma_k$ is a smoothing factor, and we define 

\begin{align}
\ell_{n,k}\left(\bdTheta;\bdTheta^{(k-1)}\right)
&=\Exp\left[\ln p\left(\bdy_n(k), \bdx_n(k),\bdx_n(k-1);\bdTheta\right)\big| \bdy^{1:k}_n,\bdTheta^{(k-1)}\right].
\end{align}
Note that the iteration index in the E-step in \eqref{online_Estep_eqn} 
is the same as the time index due to the online nature of the algorithm. Assuming a constant 
smoothing factor with $\gamma_k=1-\alpha$, where $\alpha$ can be optimized on an initial dataset \cite{Schwartz_2015}, the E-step in \eqref{online_Estep_eqn} can be written 
in compact form as 
\begin{align}\label{online_Estep_eqn2}
L_{n,k}\left(\bdTheta;\bdTheta^{(k-1)}\right)&=\alpha 
L_{n,k-1}\left(\bdTheta;\bdTheta^{(k-1)}\right)+(1-\alpha)
\ell_{n,k}\left(\bdTheta;\bdTheta^{(k-1)}\right)\nonumber\\
&=(1-\alpha)\sum^k_{j=1}\alpha^{k-j}\ell_{n,j}\left(\bdTheta;\bdTheta^{(j-1)}\right).
\end{align}
\begin{figure*}[bp]
\hrulefill
\normalsize
\setcounter{mytempeqncnt}{\value{equation}}
\setcounter{equation}{22}
\begin{align}\label{online_Estep2}
&L_{n,k}\left(\bdTheta;\bdTheta^{(k-1)}\right)
=-\frac{(1-\alpha)}{2}\sum^k_{j=1}\alpha^{k-j}\left[\ln|\bdQ|+
\ln\left[\left(\sigma^m_f\right)^2\left(\sigma^m_\theta\right)^2\right]+\text{tr}\left\{\bdQ^{-1}\left(\bdGamma^n_{j,j}
-\bdGamma^n_{j,j-1}-
\right.\right.\right.\nonumber\\
&\left.\left.\left.
\left(\bdGamma^n_{j,j-1}\right)^T+\bdGamma^n_{j-1,j-1}\right)\right\}
+\left(\sigma^m_f\right)^{-2}\left(|\hat{f}_n(j)|^2-2\hat{f}_n(j)m^f_{n,j}(j)+\gamma^n_{j,1}\right)
\right.\nonumber\\
&\left.
+\left(\sigma^m_\theta\right)^{-2}\left(|\hat{\theta}_n(j)|^2-2\hat{\theta}_n(j)m^\theta_{n,j}(j)+
\gamma^n_{j,2}\right)
\right]+\text{const.}
\end{align}
\setcounter{equation}{\value{mytempeqncnt}}
\end{figure*}

\setcounter{equation}{22}
In the M-step of the online EM algorithm, we maximize the objective function in 
\eqref{online_Estep_eqn2} with respect to $\bdTheta$ as in \eqref{Mstep_eqn}. 
\addtocounter{equation}{1}

Now we describe the computation of the auxiliary function in \eqref{online_Estep_eqn2} as follows. 
By using the conditional distributions from \eqref{state_eqn}, \eqref{obs_eqn}, and \eqref{ll_eqn}, 
this function can be computed as shown in \eqref{online_Estep2}, in which 
$m^f_{n,k}(k)$ and $m^\theta_{n,k}(k)$ are the first and second element of the 
vector $\bdm_{n,k}(k)$, respectively, and the scalars $\gamma^n_{k,1}$ and $\gamma^n_{k,2}$ define the 
$(1,1)$-th and $(2,2)$-th indexed elements of the matrix $\bdGamma^n_{k,k}$, respectively.
The correlation matrices $\bdGamma^n_{k-1,k-1}$, 
$\bdGamma^n_{k,k}$, $\bdGamma^n_{k,k-1}$, and the mean vector $\bdm_{n,k}(k)$ in \eqref{online_Estep2}, in 
iteration $k$, 
are defined in the following. To begin, the matrix $\bdGamma^n_{k-1,k-1}$ is defined by
\begin{align}\label{KS_eqn}
\bdGamma^n_{k-1,k-1}
&\triangleq\Exp\left[\bdx_n(k-1)\bdx^T_n(k-1)\big|\bdy^{1:k}_n,\bdTheta^{(k-1)}\right]\nonumber\\
&=\bdV_{n,k}(k-1)+\bdm_{n,k}(k-1)\bdm^T_{n,k}(k-1),
\end{align}
where by using the 
fixed-point Kalman smoothing equations from \cite{Sarkka_2013}, we get 
the covariance matrix $\bdV_{n,k}(k-1)$ 
as
\begin{align}\label{fp_KS_1}
\bdV_{n,k}(k-1)
&=\bdV_{n,k-1}(k-1)+\bdU^n_{k-1}\left(\bdV_{n,k}(k)-\bdV_{n,k-1}(k)\right)\left(\bdU^n_{k-1}\right)^T,
\end{align}
in which
\begin{align}\label{fp_KS_2}
\bdU^n_{k-1}&\triangleq\bdV_{n,k-1}(k-1)\left(\bdV_{n,k-1}(k)\right)^{-1},
\end{align}
and we get the mean 
$\bdm_{n,k}(k-1)$ in \eqref{KS_eqn} as 
\begin{align}
\bdm_{n,k}(k-1)
&=\bdm_{n,k-1}(k-1)+\bdU^n_{k-1}\left(\bdm_{n,k}(k)-\bdm_{n,k-1}(k)\right).
\end{align}
Similarly, the correlation matrix $\bdGamma^n_{k,k-1}$ in \eqref{online_Estep2} is defined as 
\begin{align}
\bdGamma^n_{k,k-1}
&\triangleq\Exp\left[\bdx_n(k)\bdx^T_n(k-1)\big|\bdy^{1:k}_n,\bdTheta^{(k-1)}\right]\nonumber\\
&=\bdm_{n,k}(k)\bdm^T_{n,k}(k-1)+\bdV_{n,k}(k)\left(\bdU^n_{k-1}\right)^T,
\end{align}
and the correlation matrix $\bdGamma^n_{k,k}$ in \eqref{online_Estep2} is given by using the Kalman filtering 
time-update equations as
\begin{align}\label{gamma_kk_eqn}
\bdGamma^n_{k,k}
&\triangleq\Exp\left[\bdx_n(k)\bdx^T_n(k)\big|\bdy^{1:k}_n,\bdTheta^{(k-1)}\right]\nonumber\\
&=\bdV_{n,k}(k)+\bdm_{n,k}(k)\bdm^T_{n,k}(k),
\end{align}
where $\bdV_{n,k}(k)$ and $\bdm_{n,k}(k)$ are defined in \eqref{time_upd_mV}. 
In fact, note that all the matrices in 
\eqref{KS_eqn}-\eqref{gamma_kk_eqn} 
are computed using the KF prediction-update and time-update equations defined in \eqref{pred_upd} and 
\eqref{time_upd_mV} in Section \ref{KF_section}. 

Next we compute the M-step of the online EM algorithm in which we 
maximize the auxiliary function in \eqref{online_Estep_eqn2} 
to recursively estimate $\bdTheta=\{\bdQ,\bdSigma\}$. To this end, to estimate $\bdQ$, we take the partial derivative of 
$L_{n,k}\left(\bdTheta;\bdTheta^{(k-1)}\right)$ with respect to the matrix $\bdQ$ and set it equal to zero. We 
get the estimate of $\bdQ$ in the $k$-th iteration as 
\begin{align}\label{Q_upd_1}
\bdQ^{(k)}=\frac{\sum^k_{j=1}\alpha^{k-j}\left(\bdGamma^n_{j,j}
-\bdGamma^n_{j,j-1}-\left(\bdGamma^n_{j,j-1}\right)^T+\bdGamma^n_{j-1,j-1}\right)}{\sum^k_{j=1}\alpha^{k-j}},
\end{align} 
where to get the recursive form for updating $\bdQ$, we write the numerator in \eqref{Q_upd_1} as
\begin{align}
\xi(k)&=\sum^k_{j=1}\alpha^{k-j}\left(\bdGamma^n_{j,j}
-\bdGamma^n_{j,j-1}-\left(\bdGamma^n_{j,j-1}\right)^T+\bdGamma^n_{j-1,j-1}\right),\nonumber\\
&=\alpha \xi(k-1)+\bdGamma^n_{k,k}
-\bdGamma^n_{k,k-1}-\left(\bdGamma^n_{k,k-1}\right)^T+\bdGamma^n_{k-1,k-1}
\end{align}
thus the update equation for $\bdQ$ is written as 
\begin{align}\label{Q_est_EM}
\bdQ^{(k)}=
&\frac{\left(\alpha\xi(k-1)+\bdGamma^n_{k,k}
-\bdGamma^n_{k,k-1}-\left(\bdGamma^n_{k,k-1}\right)^T+\bdGamma^n_{k-1,k-1}\right)}{\lambda_k},
\end{align}
in which $\lambda_k\triangleq\frac{1-\alpha^k}{1-\alpha}$ and we set $\xi(0)=0$. Similarly the matrix $\bdSigma$ 
can also be estimated in the M-step through computing the partial derivatives. Since it 
is a diagonal matrix as shown in \eqref{Sigma_mtx_eqn}, it can be easily estimated by recursively 
estimating its diagonal elements $\sigma^{m}_f$ and $\sigma^{m}_\theta$ as follows. To 
estimate $\sigma^{m}_f$, we compute 
\begin{align}\label{sigma_fest_EM}
\left(\sigma^{m,(k)}_f\right)^2
&=\frac{\alpha\xi^f(k-1)+\left(|\hat{f}_n(k)|^2-2\hat{f}_n(k)m^f_{n,k}(k)+\gamma^n_{k,1}\right)}{\lambda_k},
\end{align}
where $\xi^f(0)=0$, and 
to estimate $\sigma^m_\theta$, we use 
\begin{align}\label{sigma_thetaest_EM}
\left(\sigma^{m,(k)}_\theta\right)^2
&=\frac{\alpha\xi^\theta(k-1)+\left(|\hat{\theta}_n(k)|^2-2\hat{\theta}_n(k)m^\theta_{n,k}(k)+
\gamma^n_{k,2}\right)}{\lambda_k},
\end{align}
where we set $\xi^\theta(0)=0$. Thus, $\sigma^{m,(k)}_f$ and $\sigma^{m,(k)}_\theta$ together 
define the correlation matrix $\bdSigma^{(k)}$ in the $k$-th iteration. 

This completes the derivation of the online EM algorithm. For further detail, the pseudo-code 
of the resulting algorithm which combines EM, KF, and $\text{P}_\text{s}$FPC is given in Algorithm \ref{algo_2}.
\begin{algorithm}\label{algo_2}
  \footnotesize
\DontPrintSemicolon
\SetKwInput{KwPara}{Input}
\KwPara{$k=0$, define $s_n(0)=1$ 
for each 
node $n$, initialize $\bdm_{n,0}(0)$ and $\bdV_{n,0}(0)$.}

\While{convergence criterion is not met} 
{
	$k=k+1$\\
For each node $n$:
\begin{enumerate}[leftmargin=0.4cm]

\item[a)] Obtain the observation vector $\bdy_n(k)=\left[\hat{f}_n(k),\hat{\theta}_n(k)\right]^T$.\\

\begin{enumerate}[leftmargin=0.01cm]
\item[b)] Compute the prediction update step of KF.\\
\eIf{$k=1$}
{
\begin{enumerate}[leftmargin=0.01cm]
\item Run the prediction update of KF by computing \\ 
$\bdm_{n,k-1}(k)$ and $\bdV_{n,k-1}(k)$ from \eqref{pred_upd}.
\end{enumerate}
}
{
\begin{enumerate}[leftmargin=0.01cm]
\item Set $\bdm_{n,k-1}(k-1)=\left[f_n(k-1),\theta_n(k-1)\right]^T$\\ and 
compute $\bdV_{n,k-1}(k-1)$ using \eqref{V_upd_init2}.
\item Run the prediction update of KF by finding \\
$\bdm_{n,k-1}(k)$ and $\bdV_{n,k-1}(k)$ using \eqref{pred_upd}.
\end{enumerate}

}
\item[c)] Run the time update step of KF by computing \\ 
$\bdm_{n,k}(k)$ and $\bdV_{n,k}(k)$ using \eqref{time_upd_mV}.
\end{enumerate}
\item[d)] Estimate $\bdQ^{(k)}$ and $\bdSigma^{(k)}$ matrices using \eqref{Q_est_EM}-\eqref{sigma_thetaest_EM}.\\
\item[e)] For each $m\in \chi^{in}_n$, let 
$\bdm_{m,k}(k)\triangleq\left[m^f_{m,k}(k), m^\theta_{m,k}(k)\right]^T$, \\then update 
each node $n$ temporary variables by 
\begin{align}
x^f_n(k)&=\sum_{m\in\chi^{in}_n} w_{n,m} m^f_{m,k}(k)s_m(k-1)\nonumber\\
x^\theta_n(k)&=\sum_{m\in\chi^{in}_n} w_{n,m} m^\theta_{m,k}(k)s_m(k-1)\nonumber\\
s_n(k)&=\sum_{m\in\chi^{in}_n} w_{n,m} s_m(k-1),\nonumber
\end{align}
Next update $f_n(k)$ and $\theta_n(k)$ using
\begin{align}
f_n(k)&=\frac{x^f_n(k)}{s_n(k)}\nonumber\\
\theta_n(k)&=\frac{x^\theta_n(k)}{s_n(k)},\nonumber
\end{align}
\end{enumerate}
}
\KwOut{$f_n(k)$ and $\theta_n(k)$ for all $n=1,2,\ldots,N$}
\caption{EM-KF-$\text{P}_\text{s}$FPC Algorithm}
\end{algorithm}

\subsection{Discussions}\label{rem_1}
In this subsection, we begin with describing 
the initialization of the KF algorithm in each iteration, and then compute the 
computational complexity of the proposed EM-KF-$\text{P}_\text{s}$FPC algorithm. 
Note that the initialization of the EM algorithm is included in Section \ref{KF_sims} 
where the simulation results are discussed. 
\begin{itemize}
\item[a)] \textbf{{Initialization}:}
To perform the prediction update 
step of KF in the iteration $k=1$ of the EM-KF-$\text{P}_\text{s}$FPC algorithm, we define the mean vector $\bdm_{n,0}(0)$ 
and the covariance matrix $\bdV_{n,0}(0)$ of the $n$-th node 
as $\bdm_{n,0}(0)=[f_c, \pi]^T$ and 
$\bdV_{n,0}(0)
=\left[\begin{matrix}\sigma^2&0\\0&4\pi^2/12\end{matrix}\right]$.
Next since the posterior distribution on the 
state vector of node $n$ is a normal distribution and 
the means undergo a linear transformation in Step (e) of EM-KF-$\text{P}_\text{s}$FPC algorithm, 
we initialize the prediction-update step of KF in the $k>1$ iteration by defining the mean vector from the previous 
iteration as 
$\bdm_{n,k-1}(k-1)=\left[f_n(k-1),\theta_n(k-1)\right]^T$ and the 
covariance matrix $\bdV_{n,k-1}(k-1)$ as  
\begin{align}\label{V_upd_init2}
\bdV_{n,k-1}(k-1)&=\left[
\begin{matrix}
\frac{\sum_{m\in\chi^{in}_n}\eta^{k-2}_{n,m}\nu^{1,1}_{m}(k-1)}{|s_n(k-1)|^2}&
\frac{\sum_{m\in\chi^{in}_n}\eta^{k-2}_{n,m}\nu^{1,2}_{m}(k-1)}{|s_n(k-1)|^2} \\ \\
\frac{\sum_{m\in\chi^{in}_n}\eta^{k-2}_{n,m}\nu^{1,2}_{m}(k-1)}{|s_n(k-1)|^2}&
\frac{\sum_{m\in\chi^{in}_n}\eta^{k-2}_{n,m}\nu^{2,2}_{m}(k-1)}{|s_n(k-1)|^2}
\end{matrix}
\right],
\end{align}
where $\eta^{k-2}_{n,m}\triangleq|w_{n,m}s_m(k-2)|^2$ and the components $\nu^{1,1}_{m}(k-1)$, $\nu^{1,2}_{m}(k-1)$, and 
$\nu^{2,2}_{m}(k-1)$ are the $(1,1)$-th, $(1,2)$-th, and $(2,2)$-th elements, respectively, of the covariance matrix 
computed at node $m$ in the $(k-1)$-st iteration of KF using \eqref{time_upd_mV}. 
Noticeably the proposed EM-KF-$\text{P}_\text{s}$FPC algorithm is a distributed algorithm in which the nodes 
run the EM and KF algorithms on their own observations, in parallel, and then locally 
share their MMSE estimates and the error covariances 
with their neighboring nodes to update the electrical states in step $(e)$ across the array. The shared covariances are 
used to define the priors at each node for Kalman filtering in step $(b)$.
\item[b)] \textbf{{Computational Complexity}:}
For the $n$-th node in the array, the 
computational complexity of EM-KF-$\text{P}_\text{s}$FPC per iteration is dominated by \eqref{temp_PsFPC}, \eqref{KF_gain}, 
and \eqref{fp_KS_2}. Equation \eqref{temp_PsFPC} is part of the $\text{P}_\text{s}$FPC algorithm which has the 
computational complexity of $\mathcal{O}(\text{card}\{\chi^{in}_n\})$, where the notation $\text{card}\{\chi^{in}_n\}$ 
defines the cardinality of the set $\chi^{in}_n$. Equations \eqref{KF_gain} and \eqref{fp_KS_2} are part of the 
KF and EM algorithms, respectively, which require inverting and then multiplying the $2\times 2$ matrices.
These two equations have the computational complexity of $\mathcal{O}(8)$. Now since 
the KF and EM step in the synchronization algorithm 
for each node $n$ can be run in parallel, the computational complexity 
of the overall EM-KF-$\text{P}_\text{s}$FPC algorithm is $\mathcal{O}(\text{card}\{\chi^{in}_n\}+8)$. This shows 
that for the sparsely connected arrays with $\text{card}\{\chi^{in}_n\}\ll 8$, 
the computational complexity is $\mathcal{O}(8)$, whereas for the moderately connected 
large arrays with $\text{card}\{\chi^{in}_n\}\gg 8$, it can be approximated as $\mathcal{O}(\text{card}\{\chi^{in}_n\})$. 
Thus, for the moderately connected large arrays, using 
the EM and KF algorithms result in an improved synchronization performance of $\text{P}_\text{s}$FPC without any 
additional increase in the computational complexity. This result is the 
same as was observed for the KF-DFPC algorithm in \cite{rashid2022frequency} which was 
proposed for the bidirectional array networks. Furthermore, 
we also note that the computational complexity of EM-KF-$\text{P}_\text{s}$FPC is 
significantly lower than that of the VB-HCMCI algorithm in \cite{VB_HCMCI_2021} 
which not only does additionally require consensus on measurements and consensus on the 
predicted information at each node that comes at the cost of increase in the bandwidth 
requirement between the nodes and the increase in the latency for encoding the information, but they also 
need to perform multiple iterations at each time instant $k$ and at each node $n$ for the VB's convergence, before 
progressing on to the next time instant $k+1$.
\end{itemize}
\begin{figure}[tp]
    \begin{minipage}{0.48\textwidth}
        \centering
\includegraphics[width=0.99\textwidth]{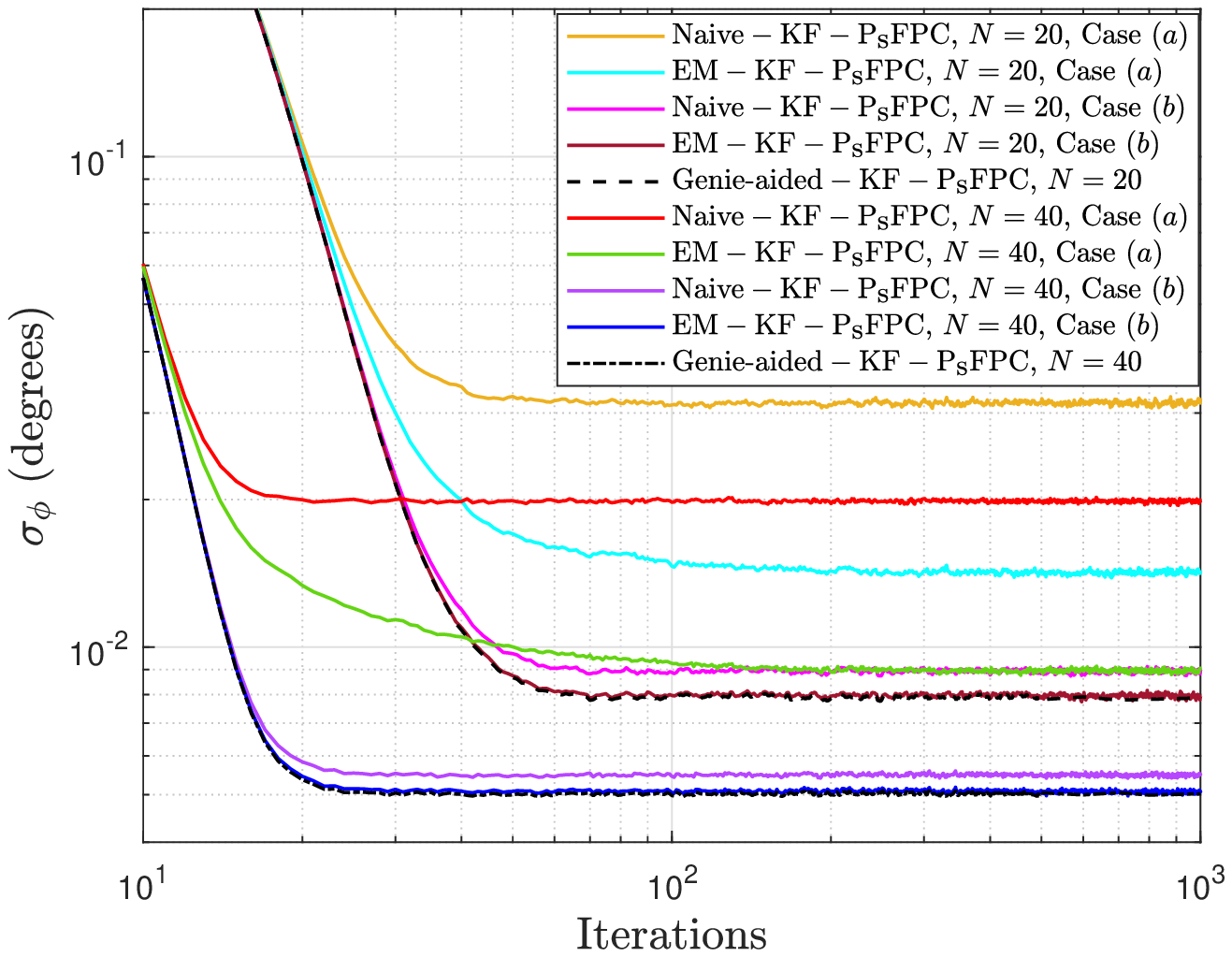}
	\caption{Standard deviation of the total phase errors of the EM-KF-$\text{P}_\text{s}$FPC and EM-KF-DFPC algorithms  vs. 
	the number of iterations for different $N$ values and for different initialization of EM when $c=0.2$, $T=0.1$ ms, and 
	SNR$=30 \text{dB}$.}
	\label{EM_KF_PsFPC_stderr_vs_Iter}			
		\end{minipage}\hspace{.025\linewidth}
    \begin{minipage}{0.48\textwidth}
        \centering
\includegraphics[width=0.99\textwidth]{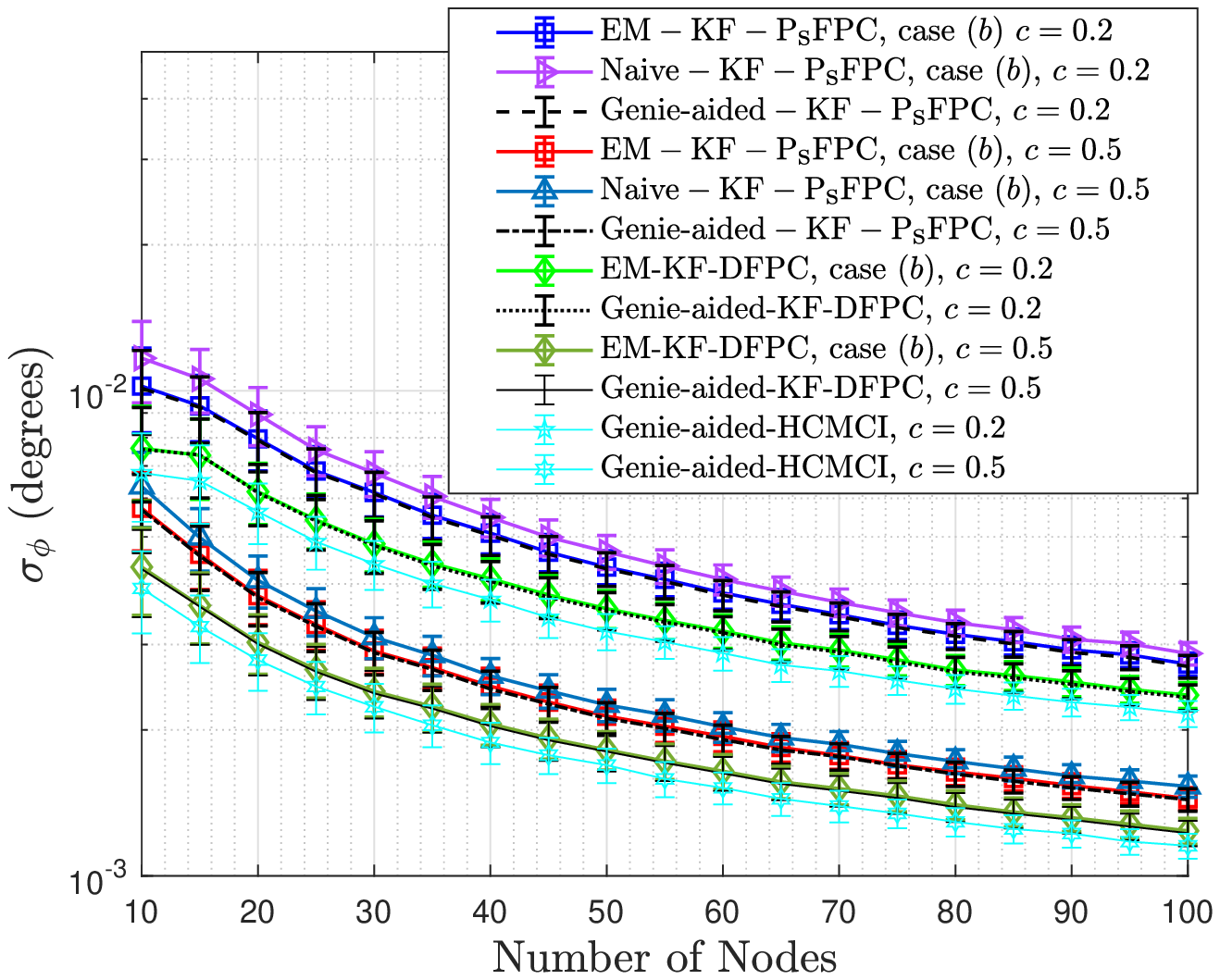}
\caption{Standard deviation of the total phase errors of the EM-KF-$\text{P}_\text{s}$FPC and EM-KF-DFPC algorithms  vs. 
	the number of nodes $N$ for different $c$ values when SNR$=30$ dB and $T=0.1$ ms.}	
	\label{EM_KF_PsFPC_stderr_vs_N}		
		\end{minipage}
\end{figure}


\subsection{Simulation Results}\label{KF_sims}

In this subsection, we evaluate the frequency and phase synchronization performances of the proposed 
EM-KF-$\text{P}_\text{s}$FPC algorithm by 
varying the number of nodes $N$ in the array and the connectivity $c$ between the nodes. To this end, we consider 
different initialization of the online EM algorithm. 
For the comparison purposes, we plot the performance of the KF-$\text{P}_\text{s}$FPC with known innovation noise 
and measurement noise covariance matrices and refer to it as the Genie-aided-KF-$\text{P}_\text{s}$FPC. In addition, 
we also show the performance of KF-$\text{P}_\text{s}$FPC when these noise covariance matrices are not updated over 
the iterations and the algorithm uses the initially estimated covariance matrices which is referred to herein as 
the Naive-KF-$\text{P}_\text{s}$FPC algorithm. 
The carrier frequency of the nodes is chosen as $f_c=1$ GHz, the sampling frequency is set as $f_s=10$ MHz, 
the update interval is $T=0.1$ ms, the SNR$=30$ dB. The results in this section were averaged over $10^3$ independent trials. 

In Fig. \ref{EM_KF_PsFPC_stderr_vs_Iter}, we analyze the convergence properties 
of the EM-KF-$\text{P}_\text{s}$FPC algorithm by evaluating the 
standard deviation of the residual phase error vs. the number of iterations for $N=20$ and $40$ nodes in the array with 
connectivity $c=0.2$, and 
for different initialization of the noise covariance matrices ($\bdQ$ and $\bdSigma$) 
in the online EM algorithm. 
For the demonstration purposes, 
we consider two cases of the initializations for the EM algorithm in EM-KF-$\text{P}_\text{s}$FPC, 
i.e., case $(a)$ represents a poor initialization of EM with  
$\bdQ^{(0)}=\left[\begin{matrix}\hat{\sigma}^2_f& 0\\0& \pi^2T^2\hat{\sigma}^2_f\end{matrix}\right]$ 
in which $\hat{\sigma}^2_f=\frac{1}{\sqrt{T}}$, and with $\bdSigma^{(0)}=\left[\begin{matrix}10^{3} &0\\
0&10^{-12}\end{matrix}\right]$, whereas case $(b)$ represents a good initialization of EM with 
$\bdQ^{(0)}=\bdQ$ and $\bdSigma^{(0)}=\left[\begin{matrix}10^{3} &0\\
0&10^{-12}\end{matrix}\right]$. 
Note that the initialization 
in case $(a)$ uses a poor estimate of $\bdQ$ that 
ignores the cross-correlation terms in $\bdQ$, uses an estimate of $\sigma^2_f$, and assumes that there is no 
phase jitter at the nodes with $\hat{\sigma}_\theta=0$. In 
contrast, the initialization in case $(b)$ uses the true $\bdQ$ matrix. Furthermore, both cases use 
a poor estimate of $\bdSigma$ matrix where the true diagonal elements can be computed using the considered 
simulation parameters in \eqref{Sigma_mtx_eqn} as $\left(\sigma^m_f\right)^2=1.52\times 10^5$ 
and $\left(\sigma^m_\theta\right)^2=4\times 10^{-12}$. 
For both of these cases and for both $N=20$ and $40$ nodes, we evaluate the 
residual phase error of EM-KF-$\text{P}_\text{s}$FPC vs. the number of iterations and compare it against the 
Naive-KF-$\text{P}_\text{s}$FPC and the 
Genie-aided-KF-$\text{P}_\text{s}$FPC as shown in the figure. It is observed that 
with the case $(a)$ initialization and for both $N$ values, the 
Naive-KF-$\text{P}_\text{s}$FPC algorithm although converges faster but results in a higher residual phase error 
as compared to EM-KF-$\text{P}_\text{s}$FPC. Our EM-KF-$\text{P}_\text{s}$FPC algorithm 
takes about $50$ to $80$ more iteration to converge when $N$ varies from $20$ to $40$ nodes, respectively, than the 
Naive-KF-$\text{P}_\text{s}$FPC algorithm 
but significantly reduces the residual phase error upon convergence. Note 
that, in this case, the residual phase error of EM-KF-$\text{P}_\text{s}$FPC is far from the 
Genie-aided-KF-$\text{P}_\text{s}$FPC for both $N$ values 
due to the convergence of EM to the local maximum near the initial estimate as discussed in Section 
\ref{EM_sectn}. In contrast, in case $(b)$, where a better 
initial estimate is available, the Naive-KF-$\text{P}_\text{s}$FPC continues to show a higher 
residual phase error as compared to EM-KF-$\text{P}_\text{s}$FPC, whereas the EM-KF-$\text{P}_\text{s}$FPC algorithm follows 
both the convergence speed and the residual phase error of the Genie-aided-KF-$\text{P}_\text{s}$FPC. 
To summarize, in both cases, EM-KF-$\text{P}_\text{s}$FPC performs better than Naive-KF-$\text{P}_\text{s}$FPC, and in 
particular with a good initialization of EM, EM-KF-$\text{P}_\text{s}$FPC converges in residual phase error 
to the Genie-aided-KF-$\text{P}_\text{s}$FPC as expected. 

In Fig. \ref{EM_KF_PsFPC_stderr_vs_N}, we continue with the case $(b)$ initialization and plot 
the final standard deviations of the 
residual phase errors of the EM-KF-$\text{P}_\text{s}$FPC, Naive-KF-$\text{P}_\text{s}$FPC, and 
Genie-aided-KF-$\text{P}_\text{s}$FPC algorithms after $100$ iterations by varying the number of nodes $N$ in the 
array and the connectivity $c$ between the nodes. It is observed that as the value of $c$ or $N$ increases, the performance 
of all the algorithms improves due to the increase in the average number of connections per node ($D$) and a decrease in 
the network algebraic connectivity ($\lambda_2$) as dicussed before. 
Moreover, by comparing Figs. \ref{fig:PsFPC_stderr_vs_c} and \ref{EM_KF_PsFPC_stderr_vs_N}, it is observed that 
Kalman filtering significantly reduces the residual phase error of the consensus-based algorithms and thereby minimizes the decoherence between the nodes.
For all the $c$ and $N$ values, the 
EM-KF-$\text{P}_\text{s}$FPC algorithm continues to perform better 
than the Naive-KF-$\text{P}_\text{s}$FPC algorithm and similar to the Genie-aided-KF-$\text{P}_\text{s}$FPC algorithm. 
For this figure, we also 
extended the KF-DFPC algorithm of \cite{rashid2022frequency} to integrate the online EM algorithm with it, and thus 
evaluated the performance of the EM-KF-DFPC algorithm and the Genie-aided-KF-DFPC algorithms. It is observed that 
for both $c=0.2$ and $0.5$, and all $N$ values, EM-KF-DFPC performs similar to Genie-aided-KF-DFPC but better than the 
EM-KF-$\text{P}_\text{s}$FPC algorithm. This is expected because the DFPC-based algorithms were evaluated using undirected 
array networks which entails more exchange of information between the nodes as compared to the directed networks which 
were used for the $\text{P}_\text{s}$FPC-based algorithms.
Note that, as mentioned before, the DFPC-based algorithms are 
only applicable to the undirected array networks with bidirectional links between the nodes, whereas 
for the directed array networks, they do not converge due 
to the use of average consensus \cite{Fast_MC_2004} as the underlying algorithm. However, our proposed 
$\text{P}_\text{s}$FPC-based algorithms are applicable for the both types of array networks, whereas in the case of 
undirected networks $\text{P}_\text{s}$FPC-based algorithms results in the same synchronization and convergence 
performance as the DFPC-based algorithms. Thus, the performances of the 
DFPC-based algorithms in this work can also be interpreted as the performances of the 
$\text{P}_\text{s}$FPC-based algorithms for the undirected array networks to drew a comparison of 
directed vs. undirected 
array network. 

Finally, Fig. \ref{EM_KF_PsFPC_stderr_vs_N} also shows the residual phase error of the HCMCI algorithm of \cite{HCMCI_2015} for the undirected array networks with $L=1$ 
consensus steps, which is 
the base algorithm of VB-HCMCI (see Table I in \cite{VB_HCMCI_2021}). 
To this end, we assume here that the noise covariance 
matrices are known to the HCMCI algorithm, and thus refer to it as the Genie-aided-HCMCI algorithm.  
It is observed that the Genie-aided-HCMCI performs slightly better than 
the EM-KF-DFPC (EM-KF-$\text{P}_\text{s}$FPC) algorithm for all the $c$ and $N$ values. As discussed earlier, this is because 
HCMCI-based algorithms additionally need to perform a consensus on the neighboring nodes' 
measurements and a consensus 
on their predicted information matrices in each iteration 
as compared to the KF-DFPC and KF-$\text{P}_\text{s}$FPC based algorithms 
which only perform consensus on the MMSE estimates and the error covariances at the neighboring nodes. Hence, the 
latter ones are computationally less complex and have 
lower latency in encoding the information and the lower bandwidth requirement for sharing that information between 
the nodes as compared 
to the HCMCI-based algorithms.


\section{Conclusions}\label{conclu}

We considered the problem of joint frequency and phase synchronization of the nodes in distributed phased arrays. 
To this end, we included the frequency drift and phase jitter of the oscillator, as well as the frequency and phase 
estimator errors in our signal model. We considered a more practical case, that is when there exist directed communication 
links between the nodes in a distributed phased array. A push-sum protocol based frequency and phase consensus algorithm referred to herein 
as the $\text{P}_\text{s}$FPC algorithm was proposed. Kalman filtering was also integrated with $\text{P}_\text{s}$FPC to 
propose the KF-$\text{P}_\text{s}$FPC algorithm. KF assumes that the innovation noise and the 
measurement noise covariance matrices are known which is an impractical assumption. An online EM based algorithm is developed that iteratively computes the ML estimates of these unknown matrices which are then used for KF. The 
residual phase error of EM upon convergence is defined by the initialization which is expected due to the non-concave structure of the objective function. However, given a good initialization, the proposed EM-KF-$\text{P}_\text{s}$FPC 
algorithm significantly improves the residual phase error as compared to the $\text{P}_\text{s}$FPC and the 
Naive-KF-$\text{P}_\text{s}$FPC algorithms, and 
also shows similar synchronization and convergence performance as that of the Genie-aided-KF-$\text{P}_\text{s}$FPC.
The residual phase error and the convergence speed of EM-KF-$\text{P}_\text{s}$FPC improves by increasing the number of nodes in the array and 
the connectivity between the nodes.  
Furthermore, the computational complexity of EM-KF-$\text{P}_\text{s}$FPC is lower than that of the HCMCI-based algorithms 
and is the same as the computational complexity of the 
KF-$\text{P}_\text{s}$FPC and KF-DFPC algorithms. In particular, for the moderately connected large array, the use of EM and KF doesn't increase the computational complexity of $\text{P}_\text{s}$FPC.

\bibliographystyle{IEEEtran}
\bibliography{References}
\end{document}